\documentclass[twoside,twocolumn,british]{revtex4-2}
\usepackage[LGR,T1]{fontenc}
\usepackage{textcomp}
\usepackage[utf8]{inputenc}
\usepackage{geometry}
\usepackage{parskip}
\usepackage{float}
\usepackage{mathrsfs}
\usepackage{amsmath}
\usepackage{amssymb}
\usepackage{graphicx}

\makeatletter

\newcommand*\LyXZeroWidthSpace{\hspace{0pt}}
\DeclareRobustCommand{\greektext}{%
  \fontencoding{LGR}\selectfont\def\encodingdefault{LGR}}
\DeclareRobustCommand{\textgreek}[1]{\leavevmode{\greektext #1}}

\usepackage{tikz}
\usetikzlibrary{arrows.meta}
\usepackage{mathrsfs}

\makeatother

\usepackage{babel}
\begin{document}
\title{The Role of Electron Correlations in Chirality Induced Spin Selectivity
of Molecular Junctions}
\author{A. Konidena}
\affiliation{Physics Department, King's College London, London, WC2R 2LS, U.K.}
\author{J. Fransson}
\affiliation{Department of Physics and Astronomy, Uppsala University, Box 516,
751 20 Uppsala, Sweden.}
\author{L. Kantorovich}
\affiliation{Physics Department, King's College London, London, WC2R 2LS, U.K.}
\begin{abstract}
The Non-Equilibrium Green's Function (NEGF) approach is applied to
a chain of Hubbard atoms in a helical geometry sandwiched between
two leads, one of which is magnetic, to assess what role the electronic
correlation effects within the molecule may play in explaining the
Chirality Induced Spin Selectivity (CISS) effect. The correlation
effects are treated at the level of the self-consistent second Born
approximation.Two chiralities, Laevorotatory $\left(\mathscr{L}\right)$
and Dextrorotatory $\left(\mathscr{D}\right)$, of the helix, are
discussed. Further, we account for spin-orbit coupling (SOC) within
the chains using a simplified model of Kane and Mele that exploits
next-nearest neighbour interactions intrinsic to a curved geometry.
The interaction with the leads is considered within the Wide Band
Approximation. We consider the ratio, $U/t$, of the Hubbard constant
$U$ to the hopping $t$ between atomic sites along the chain of up
to two, but found very little spin polarisation in our calculations.
Even though the spin polarisation factor defined as a relative difference
between the largest and smallest total current is found for both enantiomers
practically the same, the orientation of the tip's polarisation required
to reach the largest and smallest currents for the two enantiomers
differed. It is also found that the qualitative agreement with observations
can only be achieved if we account, self-consistently, for the second
order of the perturbation theory. The distribution of the spin populations
on the chain atoms for different $U/t$ ratios is also discussed.
This study concludes that even though accounting for electron-electron
interactions are important in understanding spin selectivity within
molecular junctions, they cannot be made solely responsible for the
high spin polarisation factors observed in experiments.
\end{abstract}
\maketitle

\section{Introduction}

Reports of Chirality Induced Spin Selectivity (CISS) have been ubiquitous
since its 1999 debut in Ray \textit{et. al.'s} seminal paper, `Asymmetric
Scattering of Polarised Electrons by Organised Organic Films of Chiral
Molecules' \cite{ray_asymmetric_1999}. Their original experiment
involved measurement of photoelectron transmission ejected from a
gold (Au) substrate scattered through a chiral Languir-Blodgett thin
film. Spin-selectivity manifested as asymmetry in transmission probability
when using clockwise polarised light as opposed to counter-clockwise,
while linearly polarised light was somewhere in between. Specifically,
the Laevorotatory $\left(\mathscr{L}\right)$ enantiomer yielded more
transmitted electrons from right-handed incident light than its linear
counterpart and even more so compared to left-handed polarised light,
with reversed behaviour for the Dextrorotatory $\left(\mathscr{D}\right)$
enantiomer \cite{ray_asymmetric_1999}. Thus, the introduction of
chiral molecules acting as spin-filters. Since then, however, CISS
has materialised in schemes beyond just transmission; it is used as
a broader classification for the slew of observational phenomena relating
to the effect of chiral geometry and its selection of electron spin
across various electron processes.

The revelatory benefits of spin-selective organic molecules reach
beyond the realm of biological processes and into the advancement
of memory technology by way of `spintronics', the portmanteau describing
spin transport electronics \cite{naaman_chiral_nodate}. Spin manipulation
has been championed as the method of choice for the next generation
of non-volatile, low-energy memory devices and as such has carved
a space in the spotlight for CISS \cite{thoss_perspective_2018,varela_inelastic_2014,naaman_chiral_nodate}.
With its gained popularity, CISS has had robust verification on various
molecules including double-stranded DNA of increasing lengths, helicenes
and assorted oligopeptides \cite{waldeck_spin_2021,mondal_chiral_2015,mondal_long-range_2020,lu_highly_2020}.
Some studies even isolate the geometry and compare two enantiomers
of the same compound \cite{di_nuzzo_circularly_2020,nino_enantiospecific_2014}.

These transmission studies heavily imply the persistence of CISS in
the transport of electrons through a chiral medium. Electron transport
measurements are generally conducted across a molecular junction where
an applied bias pushes electrons across the lead-molecule-lead system.
Some experiments use atomic force microscopes (AFM) \cite{lu_highly_2020,mishra_length-dependent_2020,mishra_spin-dependent_2019,mondal_long-range_2020,xie_spin_2011},
while others use a scanning tunnelling microscope (STM) \cite{aragones_measuring_2017},
depending on the system in question. Early studies used a ferromagnetic
(FM) substrate on which they grew a monolayer of organic chiral molecules.
This device provides a possibility to inject spin into the chiral
layer \cite{abendroth_analyzing_2017,mondal_long-range_2020}, where
a conductive probe AFM (CP-AFM) or a magnetic CP-AFM (mCP-AFM) were
used \cite{mishra_spin-dependent_2019,lu_highly_2020,mishra_length-dependent_2020}.

\begin{figure}
\centering
\includegraphics[width=8cm]{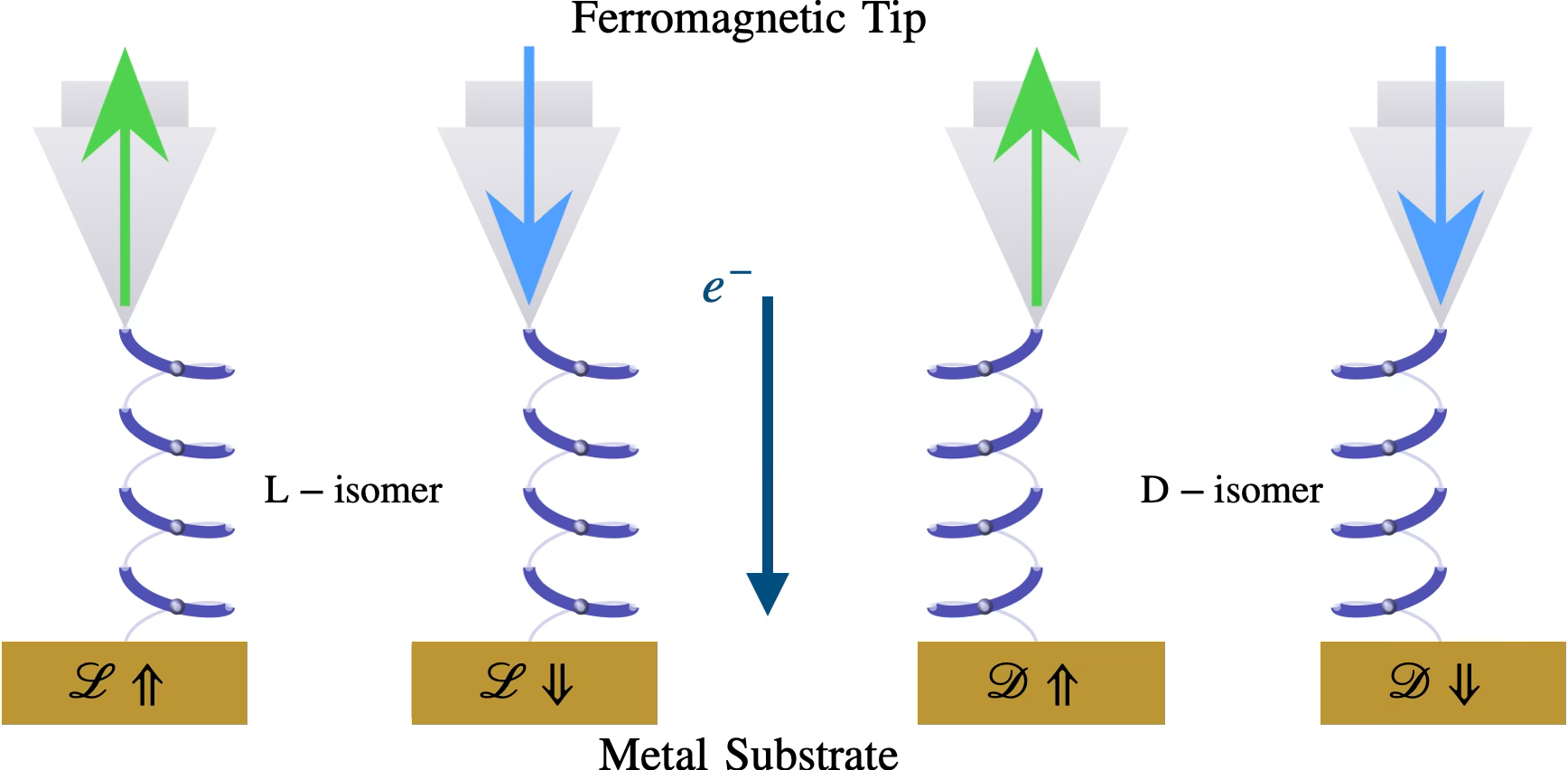}\caption{There are four possibilities to measure the current through a ferromagnetic
tip attached to a chiral molecule on a metal substrate: $\mathscr{L}$-isomer
with either up- (green arrow, $\Uparrow$) or down-polarised (blue
arrow, $\Downarrow$) tip, and $\mathscr{D}$-isomer of the chiral
molecule with both polarisations of the tip, $\Uparrow$ and $\Downarrow$.
In all four cases the chemical potential is chosen such that electrons
transmit from the tip to the surface as indicated. }\label{fig:four-cases}
\end{figure}

To observe single molecule systems, an STM is employed where the STM
tip can be either ferromagnetic (FM), e.g. nickel (Ni), or simply
conducting (e.g. Au). Using FM tips allows for electrons with one
of the spins (either up or down) to be preferentially injected into
the molecule, which could also be either of the two chiralities ($\mathscr{L}$
and $\mathscr{D}$ enantiomers). Note that in the experiment it is
not possible to measure the spin polarisation of the current directly
(i.e., currents due to electrons of either spin separately) as only
the total current due to both spins can be detected. Two polarisations
of the tip are used resulting in four total currents being measured,
see Fig. \ref{fig:four-cases}. The asymmetry (spin polarisation,
SP) factor is then defined as 
\begin{equation}
\text{SP}_{\mathscr{L}/\mathscr{D}}=\frac{J_{\mathscr{L}/\mathcal{\mathscr{D}}}^{\text{high}}-J_{\mathscr{L}/\mathscr{D}}^{\text{low}}}{J_{\mathscr{L}/\mathcal{\mathscr{D}}}^{\text{high}}+J_{\mathscr{L}/\mathscr{D}}^{\text{low}}},\label{eq:SP-def}
\end{equation}
where $J_{\mathscr{L}}^{\text{how}}$ and $J_{\mathscr{L}}^{\text{low}}$
are the highest and lowest, respectively, currents (or conductance)
measured for the $\mathscr{L}$ isomer when comparing the two tip
polarisations, and similarly for the $\mathscr{D}$ isomer \cite{aragones_measuring_2017}.
Similar (but not identical) definitions of the asymmetry factors have
been used in other work \cite{mishra_spin-dependent_2019,mishra_length-dependent_2020,huisman_chirality-induced_2023,fransson_chirality-induced_2019};
however, they capture basically the same physics. In this paper the
above definition will be employed.

These molecular junction studies of various chiral media, from carbon
nanotubes to organic chiral molecules \cite{nino_enantiospecific_2014,alam_spin_2015,abendroth_analyzing_2017,aragones_measuring_2017,mishra_length-dependent_2020,mishra_spin-dependent_2013,mishra_spin-dependent_2019,mondal_long-range_2020,xie_spin_2011,zwang_helix-dependent_2016}
found striking evidence of CISS at the single molecule level \cite{aragones_measuring_2017},
all the way to molecules of up to 12 nm \cite{mishra_length-dependent_2020},
with the SP factor increasing linearly with molecular length \cite{jia_efficient_2020,xie_spin_2011,abendroth_analyzing_2017,mishra_spin-dependent_2019}.
Aragones et. al. investigate three manifestations of asymmetry related
to conduction in this break-junction system: chirality, magnetoresistance
and what is known as the `spinterface', the interfacial dynamics of
lead/molecule interaction (coupling). To observe CISS, they used $\mathscr{L}$
and $\mathcal{\mathscr{D}}$ enantiomers of an $\alpha$-helix peptide
with 22 amino acid residues and a ferromagnetic STM tip polarised
either `up' or `down'. When comparing the four scenarios of two tip
polarisations and two chiralities, as well as a control experiment
with a non-magnetic Au tip, they were able to show convincingly that
the $\mathscr{D}$-helix preferred the `up' emitted spin, while the
$\mathscr{L}$-helix preferred `down', both exhibiting an SP of \textasciitilde 60\%
\cite{aragones_measuring_2017,aragones_magnetoresistive_2022}.

Despite extensive substantiation, the primary driver for this phenomenon
remains elusive. Qualitatively, experiment and theory agree on features
such as length dependence, where SP seems to increase with the length
of the molecule \cite{gupta_role_2025,gutierrez_modelling_2013,fransson_chirality-induced_2019},
and even temperature independence \cite{fransson_chiral_2023,guo_spin-dependent_2014,liu_chirality-driven_2021}
where the effect has been observed at both low (a few K) and room
temperatures \cite{adhikari_interplay_2023,alam_spin_2015,aragones_measuring_2017,di_nuzzo_circularly_2020,eckshtain-levi_cold_2016,lu_highly_2020,mondal_long-range_2020,ravi_magnetoresistance_2014,xie_spin_2011}.
Quantitatively, however, the discrepancy is large. While some experimentation
with organic molecules finds SP of up to 68\% \cite{mishra_spin-dependent_2019,mishra_length-dependent_2020},
simulations with realistic parameters do not observe polarisation
higher than 20\% \cite{varela_inelastic_2014} and is generally around
1\% to 4\% \cite{fransson_chirality-induced_2019,huisman_chirality-induced_2023,shitade_geometric_2020}.
While high SP was found in more complex topology \cite{zhang_highly_2023},
no theory, thus far, has been able to quantitatively reproduce the
degree of SP seen by experimentalists in ss-helial peptides, with
polarisations of only up to 15\% when extending the limits of the
input values in simulations \cite{fransson_breaking_2026}.

To probe the source of the observed SP, various suspects have been
examined theoretically, including, but not limited to, spin-orbit
coupling (SOC), electron-phonon interactions and electron-electron
interactions.

Intuitively, SOC should be the culprit for SP, however, organic molecules
that exhibit CISS tend to have very low intrinsic SOC \cite{naaman_chiral_nodate,waldeck_spin_2021}.
Further, when working in a two-terminal \textit{single channel} regime,
such as a single-strand (ss) helical molecule in a break junction
akin to those used in experiment, it was proven that SOC can be removed
completely by a SU(2) gauge transformation, which suggests that spin
filtering due solely to SOC is impossible \cite{guo_spin-selective_2012,meyer_quantum_2002}.
Despite this, Gutierrez and colleagues successfully demonstrated that
this result merely relates to a two-level atoms: they found large
SP for ssDNA in the case of multiple energy levels per atomic site,
allowing for various pathways for the electron to travel across the
system \cite{gutierrez_modelling_2013}. In this multi-orbital manifold,
the SOC cannot be gauged away; the quantum interference across multiple
transport pathways, in tandem with the helical geometry, induces a
non-Abelian SU(2) phase yielding a small but physically significant
spin-asymmetry \cite{gutierrez_modelling_2013}.

Organic molecules exhibiting CISS generally have very small intrinsic
SOC \cite{geyer_chirality-induced_nodate,gutierrez_modelling_2013,varela_inelastic_2014},
so a common way to introduce spin selectivity into theory has been
through spinterface models, where it is suggested that the coupling
of the molecule to the magnetic leads influences the molecule to act
as an orbital filter and dictates electron transmission through the
molecule to demonstrate the spin selectivity. \cite{volosniev_interplay_2021,liu_chirality-driven_2021,guo_spin-dependent_2014,dalum_theory_2019}.
Intrinsic SOC, nevertheless, cannot be completely ruled out in interacting
systems. In Refs. \cite{yeganeh_chiral_2009,ghazaryan_analytic_2020,gutierrez_modelling_2013,varela_effective_2016,varela_inelastic_2014,fransson_chirality-induced_2019,shitade_geometric_2020}
it is introduced through analytical scattering models of a helical
potential that induces SP originating from the curvature of the helices.
Hence, SOC must play some role, but it cannot be the sole cause of
CISS, so electron correlations and electron-phonon interactions may
play a decisive role here.

Modelling the effects of many-body correlations in CISS can be done
in a number of ways. For example, when treating the effects of the
electron-phonon coupling, one may use a first-principles density-matrix
Lindblad dynamics approach to consider phonon scattering and a self-consistent
SOC, as is done in Ref. \cite{gupta_role_2025}. Refs. \cite{fransson_vibrational_2020,fransson_chiral_2023}
explicitly add an electron-phonon coupling term into the Hamiltonian
of the system and reduce the time dependent vibrational problem to
a stationary one by a Lang-Firsov unitary transformation. More commonly,
model Hamiltonians are frequently called upon to incorporate electron-electron
and electron-phonon interactions explicitly into the molecular junction
\cite{fransson_chirality-induced_2019,fransson_chiral_2023,fransson_vibrational_2020,fransson_charge_2021,gupta_role_2025,huisman_chirality-induced_2023}.
Ref. \cite{savi_chirality-induced_2025} considers a minimalistic
model incorporating Hubbard interactions and non-adiabatic vibrations,
using a current constrained approach. They consider a highly correlated,
but small system, through direct numerical diagonalisation of the
Hamiltonian, and by implementing a non-equilibrium transport code,
find SP to be no greater than 10\%. Electron-electron interactions
are explored as a main cause of CISS in Refs. \cite{fransson_chirality-induced_2019,huisman_chirality-induced_2023},
but only a qualitative agreement with experiment has been achieved.
In Ref. \cite{fransson_chirality-induced_2019}, self-consistent calculations
in the Hubbard I approximation found no SP in the uncorrelated regime
but up to 4\% in the correlated case. In contrast, the authors of
Ref. \cite{huisman_chirality-induced_2023} found SP of no more than
1\% for both the Hartree-Fock (HF) mean-field (MF) and Hubbard I approximations,
also using a self-consistent non-equilibrium transport simulation.
At the heart of all these electron interaction studies lie the Non-Equilibrium
Green's Function (NEGF) method \cite{huisman_chirality-induced_2023,utsumi_spin_2020,guo_contact_2014,guo_sequence-dependent_2012,guo_spin-selective_2012,guo_spin-dependent_2014,fransson_charge_2021,varela_effective_2016,varela_inelastic_2014,fransson_chiral_2023,fransson_chirality-induced_2019},
the facilitator of quantum transport calculations.

NEGF provides a rigorous quantum mechanical framework for treating
open systems arbitrarily far from equilibrium making it particularly
well-suited for modelling charge and spin transport through molecular
junctions. Within this formalism, the system's electronic structure
is described by Green's functions that encode both the coherent propagation
of electrons through the molecule and the coupling to external reservoirs,
while self-energy terms naturally incorporate scattering processes
arising from electron-phonon coupling, electron-electron interactions,
and lead-molecule hybridisation \cite{stefanucci_nonequilibrium_2013}.
NEGF allows for a natural way to calculate the current using the Landauer-Buttiker
\cite{guo_contact_2014,gutierrez_modelling_2013,guo_sequence-dependent_2012,guo_spin-dependent_2014,guo_spin-selective_2012,varela_effective_2016,varela_inelastic_2014,utsumi_spin_2020}
(uncorrelated regime), and Mier-Wingreen \cite{fransson_chirality-induced_2019,huisman_chirality-induced_2023}
(correlated regime) formulae, providing the observables (currents
for each direction of the spin) which can be used for the comparison
with experiment. The flexibility of NEGF allows for systematic inclusion
of many-body effects and inelastic scattering channels, though the
computational cost scales rapidly with system size and, most importantly,
the level of approximation used for self-energies.

Thus, there is a clear gap in considering electron correlations as
a key mechanism of CISS. To verify if electronic correlation is essential
to understanding the nature of CISS, it is wise to ignore other possible
intrinsic effects such as the electron-phonon interaction. As is clear
from the above, the electron-electron interaction has been considered
at low levels of theory (basically, at the mean-field level); this
may at least partially explain why NEGF-based studies of CISS have
only been able to capture qualitative trends but struggle with quantitative
agreement. Therefore, one would need to move beyond these mean-field
approximation theories by explicitly and systematically considering
electron-electron interactions by incorporating higher order contributions
to the self-energy. In this work, we shall apply a self-consistent
steady state NEGF method within the Second Born Approximation and
the Hubbard Model to simulate charge and spin transport across a molecular
junction bridged by a helical atomic chain to mimic a single strand
(ss) helical molecule. In this approximation four irreducible self-energy
diagrams are considered in a self-consistent manner effectively accounting
for an infinite series of diagrams of the diagrammatic expansion with
four irreducible self-energy insertions repeated to all orders. We
shall calculate the total currents through the helical molecule of
either chirality for different biases, with and without a magnetic
tip attached to one side in order to assess the values of the SP factor
one may achieve, while accounting for SOC via the intrinsic geometry
of the helix.

The paper is organised as follows. In section 2 we outline theory
we have used. Some of the details of the calculations are moved into
Appendices. In section 3 results of the calculations are given and
discussed, while in section 4 we draw conclusions.

\section{Theory}

\subsection{Helical Molecule}

\begin{figure}
\centering
\includegraphics[width=8cm]{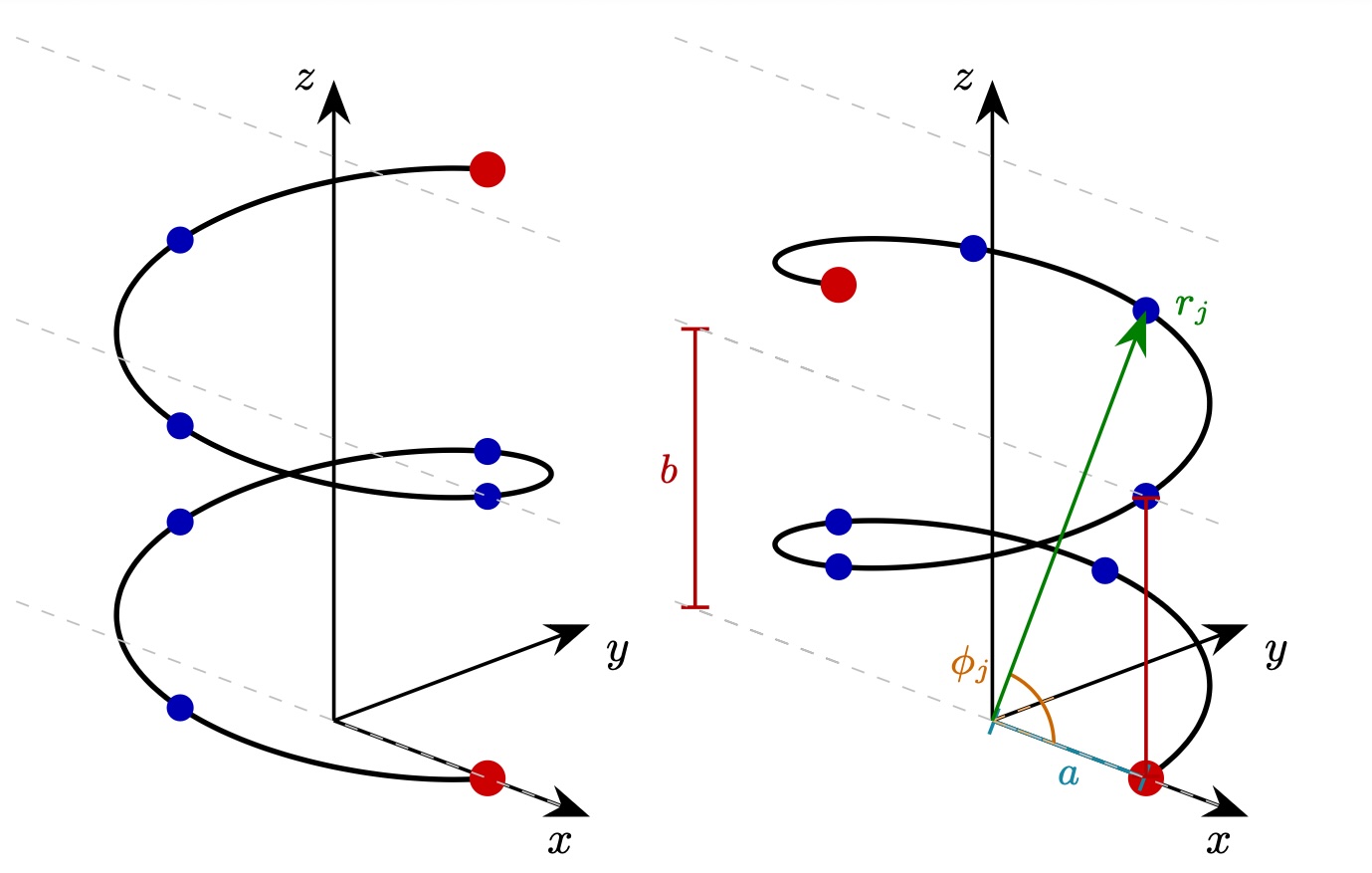}\caption{Left-handed $\left(\mathscr{L}\right)$ and right handed $\left(\mathscr{D}\right)$
helices used in our simulations containing 2 turns and 4 atomic sites
per turn (8 sites in total). The helix has radius, $a,$ pitch, $b$,
and intersite angle, $\phi$, see Eq. (\ref{eq:torsion_vector}).
The red edge atoms are the ones attached to the leads on either end
of the helix, where site 1 is attached to the tip and site 8 is attached
to the substrate (a normal metal).}\label{fig:Helix_diagram}
\end{figure}

An STM experiment such as the break junction in Ref. \cite{aragones_measuring_2017}
is modelled as a lead-molecule-lead configuration. We focus on a ss-helix
with $\mathcal{N}$ atomic sites, of radius $a,$ pitch $b$, inter-site
angle $\phi$ and atomic coordinates along the helix defined as
\begin{align}
\boldsymbol{r}_{j} & =\left(a\cos\phi_{j},a\sin\phi_{j},\frac{\left(j-1\right)bn}{\mathcal{N}-1}\right),\nonumber \\
\phi_{j} & =s\frac{n\left(j-1\right)2\pi}{\mathcal{N}},\quad j=1,...,\mathcal{N}\,,\label{eq:helix_coordinate_vec}
\end{align}
as shown in Fig. \ref{fig:Helix_diagram}, where $n$ is the total
number of turns and $s=\pm1$ corresponds to the $\mathscr{L}$ (minus)
or $\mathscr{D}$ (plus) chirality. This geometry offers a physically
naive way to introduce SOC intrinsically via the curvature of the
helix, which needs a minimum of 3 sites to be defined. One can introduce
$\boldsymbol{v}_{j}$ as a torsion vector at site $j$ and the intermediate
site as a bridge where the electron experiences the SOC. The torsion
vector is defined as 
\begin{equation}
\boldsymbol{v}_{j}=\mathbf{d}_{j+1}\times\mathbf{d}_{j+2}\,,\,\,\,\boldsymbol{d}_{j+m}=\frac{\widetilde{\mathbf{d}}_{j+m}}{\left|\widetilde{\mathbf{d}}_{j+m}\right|}\,,\label{eq:torsion_vector}
\end{equation}
where $\widetilde{\mathbf{d}}_{j+m}=\boldsymbol{r}_{j}-\boldsymbol{r}_{j+m}$.

The Hamiltonian of the molecule,
\begin{equation}
\begin{aligned}\hat{\mathcal{H}} & =\sum_{j=1}^{\mathcal{N}}\left(\sum_{\sigma=\uparrow,\downarrow}\varepsilon_{j}c_{j\sigma}^{\dag}c_{j\sigma}+U_{j}n_{j\uparrow}n_{j\downarrow}\right)\\
 & \quad-t\sum_{j=1}^{\mathcal{N}-1}\sum_{\sigma=\uparrow,\downarrow}\left(c_{j\sigma}^{\dag}c_{j+1\,\sigma}+\text{h.c.}\right)\\
 & \quad+\lambda\sum_{j=1}^{\mathcal{N}-2}\sum_{\sigma=\uparrow,\downarrow}\left(ic_{j\sigma}^{\dag}\boldsymbol{v}_{j}\cdot\boldsymbol{\sigma}c_{j+2\,\sigma}+\text{h.c.}\right)\\
 & \quad\equiv\hat{\mathcal{H}}_{C}+\sum_{j}U_{j}n_{j\uparrow}n_{j\downarrow}\,,
\end{aligned}
\label{eq:Hamiltonian}
\end{equation}

where 
\[
\hat{\mathcal{H}}_{C}=\sum_{jj'}\sum_{\sigma\sigma'}\mathcal{H}_{j\sigma,j'\sigma'}c_{j\sigma}^{\dag}c_{j'\sigma'}
\]
is the one-particle Hamiltonian of the central region that explicitly
contains the SOC. This is an adapted Kane-Mele (KM) Hamiltonian \cite{kane_quantum_2005,kane_z_2_2005},
initially adopted by Fransson for the helical geometry, employing
the nearest-neighbour hopping and next nearest neighbour SOC \cite{fransson_chirality-induced_2019,huisman_chirality-induced_2023}.
The first term describes the single site interactions with $\varepsilon_{j}$
being the on-site energy and $U_{j}$ the Hubbard parameter; $c_{j\sigma}^{\dag}$
and $c_{j\sigma}$ are the creation and annihilation operators of
an electron at site $j$ with spin $\sigma$ and $n_{j\sigma}=c_{j\sigma}^{\dagger}c_{j\sigma}$
is the corresponding occupation number operator of that spin. The
second term is responsible for the hopping between neighbouring atomic
sites through the hopping parameter $t$, while the third describes
the molecular SOC with strength $\lambda$ and the vector of three
Pauli matrices, $\sigma$.

The Hamiltonian possesses certain symmetries. Apart from being Hermitian,
$\mathcal{H}_{j\sigma,j'\sigma'}=\mathcal{H}_{j'\sigma',j\sigma}^{*}$,
it also satisfies the following conditions: $\mathcal{H}_{j\uparrow,j'\downarrow}=-\mathcal{H}_{j\downarrow,j'\uparrow}^{*}$
and $\mathcal{H}_{j\uparrow,j'\uparrow}=\mathcal{H}_{j\downarrow,j'\downarrow}^{*}$.

\subsection{Calculation of the spin currents using the Non-equilibrium Green's
Function}

\begin{figure}
\centering
\centering{}\includegraphics[width=8cm]{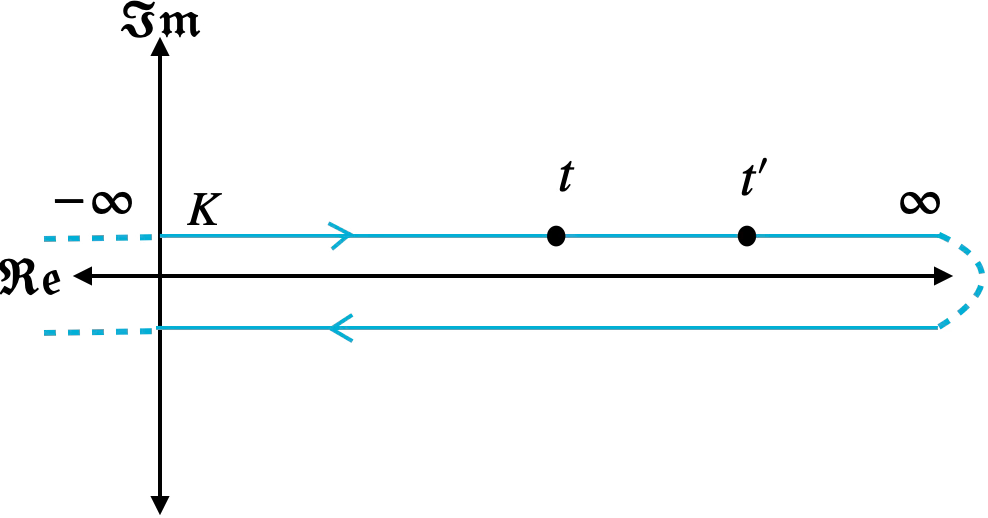}\caption{In a steady state we employ the Keldysh contour, $K$, which extends
from negative infinity to positive infinity, then curves back around
to negative infinity.}\label{fig:Steady-state-Keldysh}
\end{figure}

The NEGF \cite{stefanucci_nonequilibrium_2013,kantorovich_introduction_nodate}
is defined in the space of spin orbitals $(j\sigma)$ as
\[
G_{i\sigma,i'\sigma'}\left(z_{1},z_{2}\right)=-\frac{i}{\hbar}\left\langle \hat{T}_{K}c_{i\sigma}\left(z_{1}\right)c_{i'\sigma'}^{\dag}\left(z_{2}\right)\right\rangle ,
\]
where $z_{1}$ and $z_{2}$ are times defined on the Keldysh contour
$K$, Fig. \ref{fig:Steady-state-Keldysh}, and the operators are
defined in the Heisenberg representation; $\widehat{T}_{K}$ is the
time ordering operator on the contour. The GF forms a complex square
matrix $\mathbf{G}(z_{1},z_{2})$ of dimension $2\mathcal{N}\times2\mathcal{N}$.
We are only interested in the steady state regime here, hence the
GF when projected on its different components (retarded, advanced,
lesser, greater) would depend exclusively on the difference of real
times, $t_{1}-t_{2}$, allowing us to work in $\omega$-space by taking
the Fourier transform, as will be done henceforth.

Electron transport calculation in the steady state requires only two
independent NEGF propagators; the retarded, $\mathbf{G}^{r}(\omega)\equiv\left(G_{i\sigma,i'\sigma'}^{r}(\omega)\right)$,
and the lesser, $\mathbf{G}^{<}(\omega)\equiv\left(G_{i\sigma,i'\sigma'}^{<}(\omega)\right)$
components. The full GF is calculated by defining the bare retarded
non-interacting GF, $\mathbf{g}^{r}\left(\omega\right)=\left(g_{i\sigma,i'\sigma'}^{r}(\omega)\right)$,
of the central region using eigenvectors, $\boldsymbol{e}_{\lambda}=(e_{\lambda,i\sigma})$,
and eigenvalues, $\mathcal{E}_{\lambda}$, of the $2\mathcal{N}\times2\mathcal{N}$
Hermitian complex matrix $\boldsymbol{\mathcal{H}}=(\mathcal{H}_{i\sigma,i'\sigma'})$
of the central region Hamiltonian, $\hat{\mathcal{H}}_{C}$, which
includes the SOC. Then, 
\begin{equation}
g_{j\sigma,j'\sigma'}^{r}\left(\omega\right)=\sum_{\lambda}\frac{e_{\lambda,j\sigma}e_{\lambda,j'\sigma'}^{*}}{\hbar\omega^{+}-\mathcal{E}_{\lambda}}\,.\label{eq:g0}
\end{equation}
The noninteracting retarded GF is obtained using the following equation:

\begin{align}
\mathbf{G}_{0}^{r}(\omega)= & \left[\mathbf{g}^{r}\left(\omega\right)^{-1}-\mathbf{\Sigma}^{r,\text{emb}}\left(\omega\right)\right]^{-1}\equiv\left(\omega\mathbf{1}-\boldsymbol{\mathcal{H}}_{eff}\right)^{-1},\label{eq:G0}
\end{align}
where 
\[
\mathbf{\Sigma}^{r,\text{emb}}\left(\omega\right)=-\frac{i}{2\hbar}\sum_{\alpha}\boldsymbol{\Gamma}^{\alpha}
\]
is the sum of the left and right leads self-energies, where $\alpha=R,L$
is the lead index. The two leads and their interactions with the molecule
are described within the wide band approximation (WBA) using the level
width matrix $\boldsymbol{\Gamma}^{(\alpha)}=\left(\Gamma_{j\sigma,j'\sigma'}^{(\alpha)}\right)$
which does not depend on $\omega$. In this case the junction is described
by the effective Hamiltonian matrix 
\begin{equation}
\boldsymbol{\mathcal{H}}_{eff}=\boldsymbol{\mathcal{H}}-\frac{i}{2\hbar}\boldsymbol{\Gamma},\label{eq:Eff_hamiltonian}
\end{equation}
where $\boldsymbol{\Gamma}=\sum_{\alpha}\boldsymbol{\Gamma}^{\alpha}=\boldsymbol{\Gamma}^{R}+\boldsymbol{\Gamma}^{L}$
is the total level width function \cite{stefanucci_nonequilibrium_2013,kantorovich_introduction_nodate}.
The non-interacting lesser GF is calculated by means of the Keldysh
equation
\begin{equation}
\mathbf{G}_{0}^{<}(\omega)=\mathbf{G}_{0}^{r}(\omega)\boldsymbol{\Sigma}^{<,\text{emb}}(\omega)\mathbf{G}_{0}^{a}(\omega)\,,\label{eq:Keldysh}
\end{equation}
where the lesser self-energy due to the leads is given by 
\begin{equation}
\begin{aligned}\boldsymbol{\Sigma}^{<,\text{emb}}(\omega) & =\sum_{\alpha}\boldsymbol{\Sigma}^{<,\text{emb},\alpha}(\omega)=\frac{i}{\hbar}\sum_{\alpha}n_{F}\left(\hbar\omega_{\alpha}\right)\boldsymbol{\Gamma}^{\alpha}\,,\end{aligned}
\label{eq:lesser-emb-SE}
\end{equation}

$n_{F}\left(\hbar\omega\right)$ is the Fermi function, $\hbar\omega_{L}=\hbar\omega+eV$
and $\hbar\omega_{R}=\hbar\omega$. Here $V$ is the bias (considered
as applied to the left lead, i.e., to the tip). The remaining two
projections, the advanced and the greater, are related to the other
two via $\mathbf{G}_{0}^{a}(\omega)=\mathbf{G}_{0}^{r}(\omega)^{\dagger}$
and 
\begin{equation}
\mathbf{G}_{0}^{>}(\omega)=\mathbf{G}_{0}^{<}(\omega)+\mathbf{G}_{0}^{r}(\omega)-\mathbf{G}_{0}^{a}(\omega)\,.\label{eq:Greater}
\end{equation}

\begin{figure}
\centering
\centering{}\includegraphics[width=8cm]{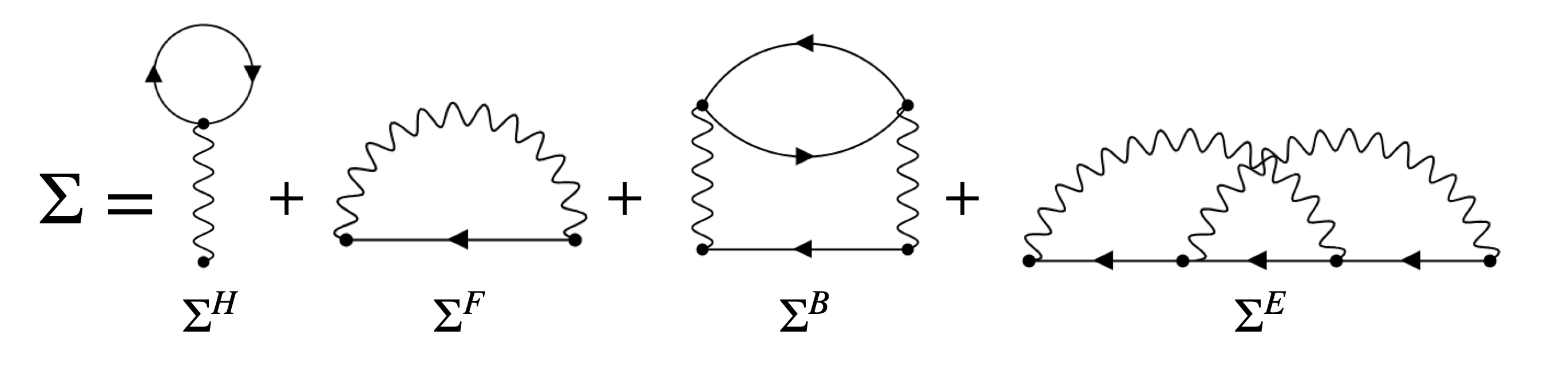}\caption{The four diagrams that contribute to SCF calculations in the SBA.}\label{fig:SBA_diagrams_noindex}
\end{figure}

Once all four components of the GF in the zero order approximation
are calculated, we can proceed to include the Hubbard interaction.
We use the interaction self-energy within the Second Born Approximation
(SBA) by taking all contributing electron-electron interaction diagrams
up to and including the second order diagrams in perturbation theory
expressed via the various total NEGF $\left(\mathbf{G}(\omega)\right)$,
components, making the calculation self-consistent. There are only
four diagrams to be included shown in Fig. \ref{fig:SBA_diagrams_noindex}:
the first two are the Hartree (H) and Fock (F) diagrams, while the
latter are the bubble (B) and exchange (E) diagrams, respectively,
$\boldsymbol{\Sigma}=\boldsymbol{\Sigma}^{H}+\boldsymbol{\Sigma}^{F}+\boldsymbol{\Sigma}^{B}+\boldsymbol{\Sigma}^{E}$.
The explicit expressions for all diagrams for both retarded and lesser
projections for our Hubbard chain of atoms, with spin being taken
into account explicitly, are given in the Appendix. Importantly, in
the first order (the Hartree and Fock diagrams) the self-energy is
the same for all $\omega$ providing only a static contribution to
the self-energy due to electronic correlations, while in the second
order (the bubble and exchange diagrams) the self-energy contributions
are \textit{dynamic} as being strongly $\omega-$dependent. Therefore,
we not only include static, but also dynamic, correlation effects
in our simulations.
 The retarded and lesser GFs with full account of the interaction
within our model are calculated from the Dyson,
\begin{align}
\boldsymbol{G}^{r}(\omega) & =\left(\mathbf{g}^{r}\left(\omega\right)^{-1}-\widetilde{\boldsymbol{\Sigma}}^{r}(\omega)\right)^{-1}\nonumber \\
 & \equiv\left(\omega\mathbf{1}-\boldsymbol{\mathcal{H}}_{eff}-\boldsymbol{\Sigma}^{r}(\omega)\right)^{-1},\label{eq:Dyson_equation}
\end{align}
and the Keldysh equations,
\begin{equation}
\boldsymbol{G}^{<}(\omega)=\boldsymbol{G}^{r}(\omega)\widetilde{\boldsymbol{\Sigma}}^{<}(\omega)\boldsymbol{G}^{a}(\omega),\label{eq:Keldysh_eqn}
\end{equation}
respectively. Here $\widetilde{\boldsymbol{\Sigma}}^{r/<}=\boldsymbol{\Sigma}^{r/<}+\boldsymbol{\Sigma}^{r/<,\text{emb}}$
are the \textit{full }self-energies (retarded and lesser) that include
the embedding part as well; note that they are functionals of the
\textit{full} retarded and lesser GFs. Therefore, to calculate both
components of the GF, we perform self-consistent field (SCF) calculations
of the Dyson and Keldysh equations. The approach we take here is explained
in more detail in Sect. \ref{subsec:Self-Consistent-Field-Calculatio}.
It is essential to stress that the effective Hamiltonian is the only
quantity that enters both the Dyson and Keldysh equations, and hence
eventually defines the GFs after solving these equations. Thus, any
symmetry inherent to the effective Hamiltonian matrix, $\boldsymbol{\mathcal{H}}_{eff}$,
will be reflected in the both retarded and lesser GFs.

Having found the converged GF's we use the Meir-Wingreen formula for
the current in the $\alpha$ lead of spin $\sigma$:
\begin{align}
J_{\sigma}^{\left(\alpha\right)}\left(\omega\right) & =\int\frac{d\omega}{2\pi}\sum_{i}\left[\Sigma_{i\sigma,i\sigma}^{>,\mathrm{emb},\alpha}\left(\omega\right)G_{i\sigma,i\sigma}^{<}\left(\omega\right)\right.\nonumber \\
 & \left.-\Sigma_{i\sigma,i\sigma}^{<,\mathrm{emb},\alpha}\left(\omega\right)G_{i\sigma,i\sigma}^{>}\left(\omega\right)\right],
\end{align}
where the trace is taken over all orbitals of the central region of
spin $\sigma$ (the molecule), and the lesser embedding self-energy
is defined by Eq. (\ref{eq:lesser-emb-SE}), while the greater one
is similarly given by 
\begin{equation}
\begin{aligned}\boldsymbol{\Sigma}^{>,\text{emb}}(\omega) & =\sum_{\alpha}\boldsymbol{\Sigma}^{>,\text{emb},\alpha}(\omega)\\
 & =\frac{i}{\hbar}\sum_{\alpha}\left[n_{F}\left(\hbar\omega_{\alpha}\right)-1\right]\boldsymbol{\Gamma}^{\alpha}\,.
\end{aligned}
\label{eq:greater-emb-SE}
\end{equation}
The total charge current is recovered by summing the two spin currents,
$J^{(\alpha)}=J_{\uparrow}^{(\alpha)}+J_{\downarrow}^{(\alpha)}.$
In practice, we calculate the current in the left lead (in the tip)
and will drop the superscript $(\alpha)$. As the electrons are injected
from the tip, a negative current is expected.

Using the converged lesser GF component allows the calculations of
site populations, $n_{i}=n_{i\uparrow}+n_{i\downarrow}$, as well
as spin populations, $s_{i}=n_{i\uparrow}-n_{i\downarrow}$, at lattice
sites via

\[
n_{i\sigma}=-i\hbar\int\frac{d\omega}{2\pi}G_{i\sigma,i\sigma}^{<}(\omega)\,.
\]

\subsection{Computational Procedure}\label{subsec:Self-Consistent-Field-Calculatio}

The interacting GFs are obtained through an iterative SCF procedure.
The process is initialised by calculating the non-interacting retarded
Green’s function, $\boldsymbol{g}^{r}$ of the central region via
Eq. (\ref{eq:g0}). Then $\boldsymbol{G}_{0}^{r}$ is calculated via
Eq. (\ref{eq:G0}), which allows us to calculate $\boldsymbol{G}_{0}^{<}$
incorporating the leads self-energies via Eq. (\ref{eq:Keldysh}),
i.e., while initially neglecting electron interactions within the
central molecular region. Using these initial propagators, we evaluate
the interaction self-energies $\Sigma$ corresponding to the diagrammatic
expansion within the SBA as illustrated in Fig. \ref{fig:SBA_diagrams_noindex}.
The interacting Green’s functions, $\boldsymbol{G}^{r}$ and $\boldsymbol{G}^{<}$,
are then determined by solving the Dyson and Keldysh equations, Eq.
(\ref{eq:Dyson_equation}) and (\ref{eq:Keldysh_eqn}), respectively.
At each subsequent iteration, the self-energies are updated using
the Green’s functions obtained from the previous step. This cycle
continues until the maximum difference between successive iterations
for both functions falls below a convergence threshold of $10^{-6}$.

The described procedure, however, does not always converge. To mitigate
this problem, the output, i.e. the retarded and lesser GF for the
$(n+1)$-th iteration is defined by a linear Pulay mixture \cite{pulay_convergence_nodate}
of the input and the output from the previous iteration:
\begin{equation}
G_{out\,(n+1)}^{r/<}=\left(1-p\right)G_{in\,(n+1)}^{r/<}+p\,G_{out\,(n)}^{r/<}\,,\label{eq:mixing_GF}
\end{equation}
where $p$ is the Pulay mixing coefficient, chosen to be between 0
and 1 as needed. This way one maintains numerical stability against
the inherent non-linearities of the Hubbard model, dampening potential
oscillations to facilitate convergence (which has proven to be difficult
at higher $U/t$ ratios ($U/t$ > 2) and longer helical chains ($\mathcal{N}$
> 8)).

\section{Results}

As seen in Fig. \ref{fig:Helix_diagram}, we model the \textgreek{α}-helix
as a tight-binding chain of $\mathcal{N}=8$ sites arranged over 2
helical turns, with a pitch and the radius of 2.5 Å and 1.0 Å, respectively
to approximate the structural characteristics of an $\alpha$-helix,
which has on average 3.54 residues per turn, a pitch of 5.4 Å and
a radius of 2.3 Å \cite{barlow_helix_1988}. The molecule is attached
on one side to a normal metal (the surface) and on the other - to
the tip, which we shall assume to be FM.

To simulate the conducting Hubbard regime appropriate for a peptide
backbone, we use an on-site energy $\epsilon=2.0$ eV and the hopping
parameter $t=0.4$ eV, placing the system well away from half-filling,
consistent with the experimentally observed conducting behaviour of
$\alpha$-helical peptides in STM junctions \cite{aragones_measuring_2017,gutierrez_modelling_2013}.
It is found that the eigenvalues of the Hamiltonian of the molecule
lie within the interval between about 1.00 and 3.00 eV for the values
of $\lambda$ between 0.05 and 0.5, hence, by choosing the chemical
potential $\mu=0$ we ensure that conductance happens at bias voltages
(measured with respect to the tip) between $-3.0$ and $-1.0$ V and
the electrons are injected from the FM tip through the molecule and
into the surface, see Fig. \ref{fig:four-cases}.

The coupling of the molecule to the leads is controlled by the level-width
matrices $\boldsymbol{\Gamma}^{\alpha}$. In this study we assume
that non-zero elements of the level-width matrices are the only ones
that correspond to the edge atoms of the molecule and which are diagonal
with respect to the spin, $\Gamma_{1\sigma,1\sigma}^{L}\ne0$ and
$\Gamma_{\mathcal{N}\sigma,\mathcal{N}\sigma}^{R}\ne0$. Moreover,
the elements $\Gamma_{\mathcal{N}\sigma,\mathcal{N}\sigma}^{R}=\Gamma^{R}$
are assumed spin-independent as corresponding to the coupling to the
normal metal, while the magnetisation state of the STM tip is encoded
in the asymmetry of the spin-resolved left-lead couplings: $\Gamma_{1\uparrow,1\uparrow}^{L}\equiv\Gamma_{\uparrow\uparrow}^{L}\ne\Gamma_{1\downarrow,1\downarrow}^{L}\equiv\Gamma_{\downarrow\downarrow}^{L}$.
The up polarisation of the FM tip ($\Uparrow$) is modelled by choosing
$\Gamma_{\uparrow\uparrow}^{L}>\Gamma_{\downarrow\downarrow}^{L}$,
while the its down polarisation ($\Downarrow$) - by choosing $\Gamma_{\downarrow\downarrow}^{L}>\Gamma_{\uparrow\uparrow}^{L}$.
When comparing the currents calculated for opposite polarisations
of the tip, we would choose the swapped values of $\Gamma_{\uparrow\uparrow}^{L}$
and $\Gamma_{\downarrow\downarrow}^{L}$. In other words, the $\Uparrow$
tip polarisation is modelled using small values of $\Gamma_{\downarrow\downarrow}^{L}$
and $\Gamma^{R}$, and a larger value of $\Gamma_{\uparrow\uparrow}^{L}$,
while the case of the tip polarisation $\Downarrow$ is modelled with
relatively small $\Gamma_{\uparrow\uparrow}^{L}$, $\Gamma^{R}$ and
a larger value of $\Gamma_{\downarrow\downarrow}^{L}$.

In this case the effective Hamiltonians, $\boldsymbol{\mathcal{H}}_{eff}$
Eq. (\ref{eq:Eff_hamiltonian}), of both chiralities, $\mathscr{L}$
and $\mathscr{D}$, and two polarisations of the tip, $\Uparrow$
and $\Downarrow$, are related to each other as follows ($\sigma$
and $\overline{\sigma}$ correspond to opposite directions of the
spin): 
\begin{align}
\left(\boldsymbol{\mathcal{H}}_{eff}\right)_{j\sigma,j'\sigma}^{\mathscr{D}\Uparrow} & =\left(\boldsymbol{\mathcal{H}}_{eff}\right)_{j\overline{\sigma},j'\overline{\sigma}}^{\mathscr{L}\Downarrow}\,\,\,\label{eq:symmetry1}\\
 & \text{and}\,\,\,\left(\boldsymbol{\mathcal{H}}_{eff}\right)_{j\sigma,j'\overline{\sigma}}^{\mathscr{D}\Uparrow}=-\left(\boldsymbol{\mathcal{H}}_{eff}\right)_{j\overline{\sigma},j'\sigma}^{\mathscr{L}\Downarrow},\nonumber 
\end{align}
as well as 
\begin{align}
\left(\boldsymbol{\mathcal{H}}_{eff}\right)_{j\sigma,j'\sigma}^{\mathscr{D}\Downarrow} & =\left(\boldsymbol{\mathcal{H}}_{eff}\right)_{j\overline{\sigma},j'\overline{\sigma}}^{\mathscr{L}\Uparrow}\,\,\,\,.\label{eq:symmetry2}\\
 & \text{and}\,\,\,\left(\boldsymbol{\mathcal{H}}_{eff}\right)_{j\sigma,j'\overline{\sigma}}^{\mathscr{D}\Downarrow}=-\left(\boldsymbol{\mathcal{H}}_{eff}\right)_{j\overline{\sigma},j'\sigma}^{\mathscr{L}\Uparrow}\nonumber 
\end{align}
As noted above, this symmetry will be reflected in the GFs and eventually
results in the following relationships between the spin currents of
the four systems: $J_{\uparrow}^{\mathscr{D}\Uparrow}=J_{\downarrow}^{\mathscr{L}\Downarrow}$,
$J_{\downarrow}^{\mathscr{D}\Uparrow}=J_{\uparrow}^{\mathscr{L}\Downarrow}$,
and $J_{\uparrow}^{\mathscr{D}\Downarrow}=J_{\downarrow}^{\mathscr{L}\Uparrow}$,
$J_{\downarrow}^{\mathscr{D}\Downarrow}=J_{\uparrow}^{\mathscr{L}\Uparrow}$.
Therefore, within our model, only two different total currents are
expected within the four systems to be studied, as the total currents
of $\mathscr{D}\Uparrow$ and $\mathscr{L\Downarrow}$, as well of
$\mathscr{D}\Downarrow$ and $\mathscr{L\Uparrow}$, are the same.
Correspondingly, in what follows, we shall only report the currents
for the two systems, $\mathscr{D}\Uparrow$ and $\mathscr{D}\Downarrow$
.

\begin{figure}
\centering
\includegraphics[width=8cm]{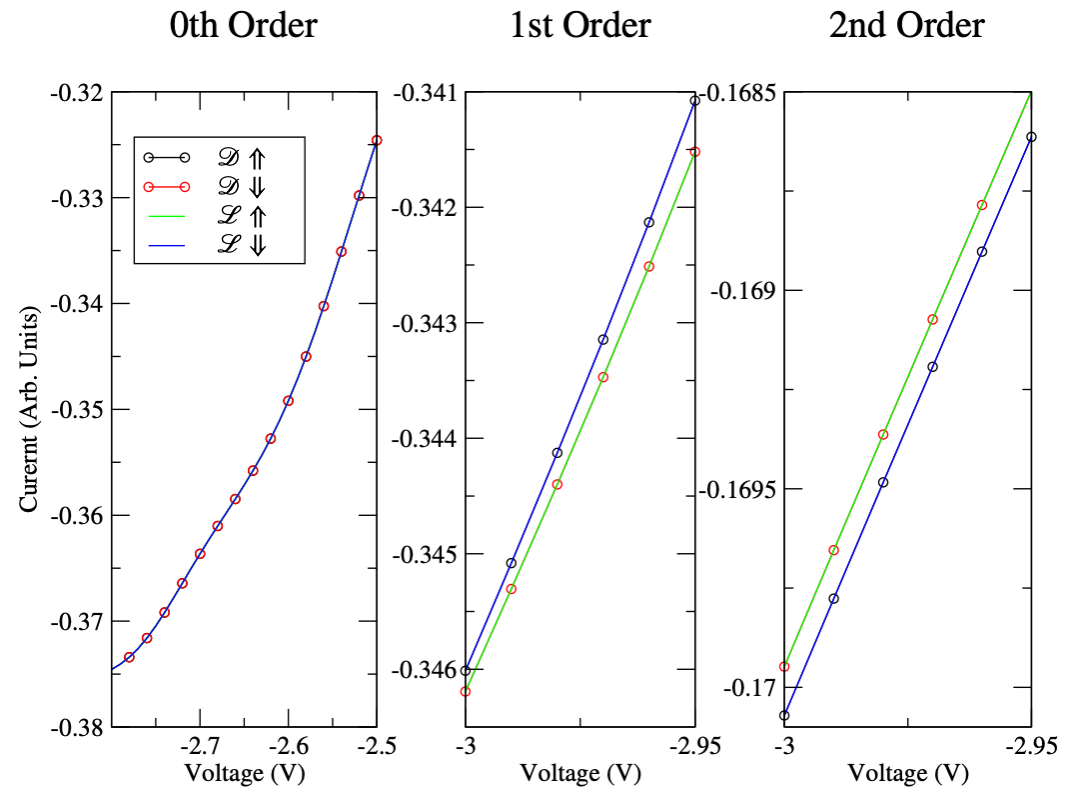}\caption{The total current for both chiralities, $\mathscr{D}$ and $\mathscr{L}$,
both tip magnetisations, $\Uparrow$ and $\Downarrow$, and all three
orders of the electron correlations. The following parameters are
used: $\varepsilon=2.0$, $t=-0.4$, $\lambda=0.05$ and $U=0.8$;
$\Gamma_{\uparrow\uparrow}^{L}=1.5$ and $\Gamma_{\downarrow\downarrow}^{L}=1.0$
for the tip polarisation $\Uparrow$, while the swapped values for
the $\Downarrow$ polarisation; in either case, $\Gamma^{R}=1.0$.}\label{fig:total}
\end{figure}

We shall first discuss a set of simulations in which $U/t=2$ and
the spin-orbit coupling strength $\lambda=t/8$. The role of the perturbation
order in the electron-electron self-energy on the total currents is
demonstrated in Fig. \ref{fig:total}, where in the zero-th order
no electron correlation diagrams are accounted for and in the first
only the Hartree-Fock diagrams are included (first two diagrams in
Fig. \ref{fig:SBA_diagrams_noindex}), while in the second - the Hartree,
Fock, Bubble and Exchange diagrams are taken into the interacting
self-energy (all four diagrams in Fig. \ref{fig:SBA_diagrams_noindex}).

The first observation is that, for all three levels of theory, the
total currents coincide for systems $\mathscr{D}\Uparrow$ and $\mathscr{L}\Downarrow$,
as well as for $\mathscr{D}\Downarrow$ and $\mathscr{L}\Uparrow$,
as mentioned above, because of the inherent symmetry of the effective
Hamiltonian. Effectively, only two distinct values of the current
are observed instead of four: for $\mathscr{D}\Uparrow$ and $\mathscr{L}\Uparrow$
($\mathscr{D}\Downarrow$ and $\mathscr{L}\Downarrow$). In that sense,
the effect of chirality is evident: for the same tip polarisation,
the currents for both chiralities indeed differ, in essence manifesting
the CISS effect. When comparing this to the experimental observation
that the $\mathscr{D}$-isomer prefers tip polarisation $\Uparrow$
and $\mathscr{L}$ prefers $\Downarrow$, we notice that this is only
reflected in the second order calculations of the total current, while
the first order presents the opposite. Secondly, the currents are
negative, consistent with the electron flow from the tip to the substrate
under the applied negative bias (defined with respect to the tip).
Thirdly, the electron-electron interactions systematically suppress
the overall current magnitude relative to the 0th order baseline $J-V$
curves across the entire voltage range, with the first- and second-order
corrections progressively pulling the curves closer toward zero.

The picture becomes even more revealing if we consider the total current
for all four systems (Fig. \ref{fig:total}). At zero-th order, the
four curves are practically indistinguishable: without electron-electron
interactions, the total current carries no imprint of either the molecular
chirality or the tip magnetisation, and the CISS signal is absent;
there is only a modest spin current splitting (not shown) arising
purely from the SOC-induced asymmetry in the transmission function
in the presence of the asymmetric tip coupling. Note, however, that
the spin currents cannot be measured separately in experiment, only
the total current can be observed. The observed indistinguishability
of the total currents is not merely a quantitative statement but a
qualitative one: the non-interacting system with only SOC and spin-asymmetric
tip coupling is insufficient to lift the degeneracy between the four
scenarios in the total current.

\begin{figure}
\centering
\includegraphics[width=8cm]{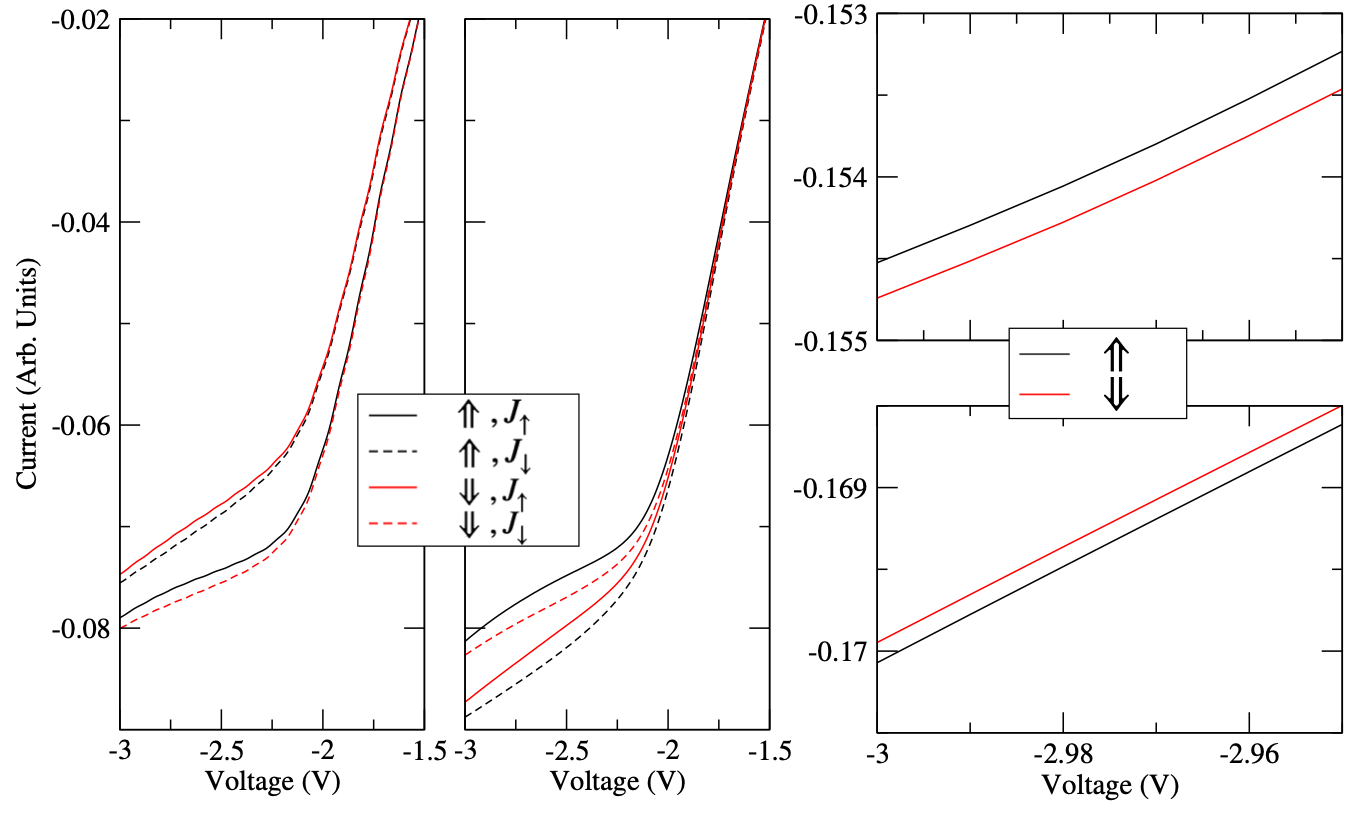}

\caption{Spin resolved currents $\left(J_{\uparrow/\downarrow}\right)$ and
the total currents as functions of the voltage for $U/t=2$, calculated
in the second order of perturbation theory for the self-energy. Left
and middle panels: spin-resolved currents for $\mathscr{D}$ helicity
with the tip polarised up $\left(\Uparrow\right)$ and down $(\Downarrow)$.
In the left panel (regime A): $\Gamma_{\uparrow\uparrow}^{L}=0.6$
and $\Gamma_{\downarrow\downarrow}^{L}=\Gamma^{R}=0.4$, while in
the middle panel (regime B): $\Gamma_{\uparrow\uparrow}^{L}=1.5$
and $\Gamma_{\downarrow\downarrow}^{L}=\Gamma^{R}=1.0$. Two right
panels: the total currents for both regimes and both tip polarisations.
The following parameters were used: $\varepsilon=2.0$, $t=-0.4$,
$\lambda=0.05$ and $U=0.8$.}\label{fig:tip_up_FS}
\end{figure}

At first and second orders, by contrast, the four curves split into
two groups of curves which are cleanly resolved. At first order, the
spin currents splitting (not shown) is markedly enhanced and remains
robust across the full voltage range demonstrating that the static
Hartree-Fock self-energy amplifies the SOC-induced asymmetry through
the spin-dependent level renormalisation provided by the Fock diagram
(see Appendix). At the second order the bubble and exchange diagrams,
which introduce dynamical correlations, partially screen out the spin
selectivity somewhat reducing the spin splitting, especially at large
$V$ (left and middle panels in Fig. \ref{fig:tip_up_FS}).

\begin{figure}
\centering
\includegraphics[width=8cm]{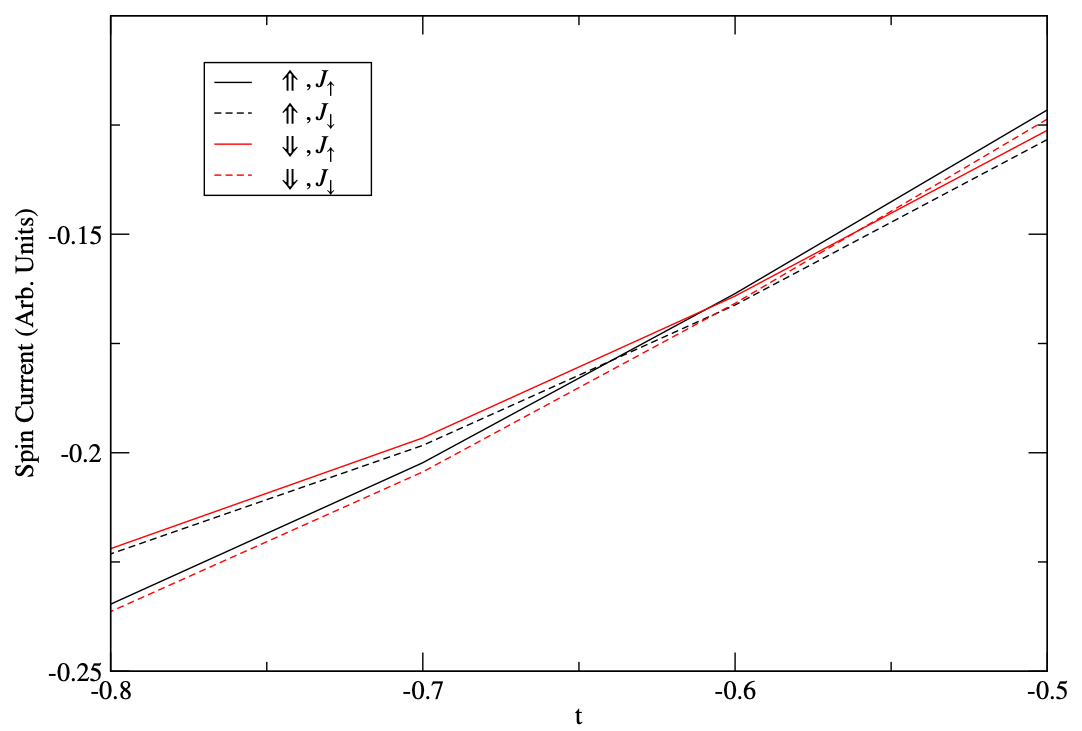}\caption{Spin-resolved currents $J_{\uparrow}$ and $J_{\downarrow}$ as a
function of the hopping parameter $t$, for two tip polarisations.
Black lines correspond to tip polarised up ($\Uparrow$) and red lines
to tip polarised down ($\Downarrow$), with solid (dashed) lines denoting
the spin-up (spin-down) current component. The level width functions
are set to $\Gamma_{\uparrow\uparrow}^{L}=1.5$ and $\Gamma_{\downarrow\downarrow}^{L}=\Gamma^{R}=1.0$
for the tip polarised up case, with the spin assignments reversed
for tip polarised down. A crossover occurs at $t\approx-0.6$, below
which the dominant current corresponds to the same spin channel as
the tip polarisation (spin-filtering regime), and above which the
dominant current corresponds to the opposite spin channel (spin-flipping
regime).}\label{fig:t_comparison}
\end{figure}

To probe the switching of the selected spin, first observed in Fig.
\ref{fig:total}, we consider our junctions for different values of
the level width functions, and have discovered two different regimes:
(A) spin filtering and (B) spin flipping. The first regime is observed
when the level width functions $\Gamma_{\uparrow\uparrow}^{L}$, $\Gamma_{\downarrow\downarrow}^{L}$
and $\Gamma^{R}$ are of the same order of magnitude as the hopping
parameter, $t$, while in the other the level width functions are
larger than $t$. The names of the two regimes given above will become
apparent when we focus on the spin-resolved currents for two choices
of the level width functions, as shown in the left and middle panels
of Fig. \ref{fig:tip_up_FS}. For the first case, the left panel,
the largest (in the absolute value) spin currents coincide with the
direction of the tip polarisation. In this case the tip injects more
electrons with spin up (down) if it is polarised up (down), and these
electrons propagate through the molecule maintaining their spin (filtering).
In the second regime (the middle panel), the situation changes qualitatively.
We observe that the largest spin currents correspond to the \textit{opposite}
direction of the tip polarisation: if the tip is polarised up (down),
the largest is the spin current $J_{\downarrow}$ ($J_{\uparrow}$).
In other words, the tip injects more electrons of a certain spin,
but then the spin of these electrons flips when passing through the
chiral molecule resulting in the largest spin-up current for the tip
polarisation $\Downarrow$, and vice versa for the tip polarisation
$\Uparrow$. The transition between these two regimes can be understood
by examining the competition between the coupling strength and the
molecular hopping. To illustrate this, Fig. \ref{fig:t_comparison}
shows the four spin-resolved currents as a function of $t$ for the
D-enantiomer with both up and down tip polarisations. At large $t$,
where the hopping parameter is of comparable magnitude to the \textgreek{Γ}'s,
the junction operates in the spin-filtering regime: the spin currents
are ordered according to the tip polarisation, with the dominant current
flowing in the same spin channel as the injected electrons. As $t$
decreases below the crossover point (t \ensuremath{\approx} \textminus 0.6),
the relative ordering of the spin currents inverts, signalling the
transition into the spin-flipping regime. In this limit, the strong
coupling to the leads relative to the intramolecular hopping seems
to endorse a spin flipping mechanism across the chiral molecule.

This observation is also supported by the calculation in which instead
we change the injection strength, $\Gamma_{\uparrow\uparrow}^{L}$,
for one spin direction. There is a crossing of the two spin currents,
$J_{\uparrow}$ and $J_{\downarrow}$, at around $\Gamma_{\uparrow\uparrow}^{L}\approx0.5$,
signifying that after this point for larger values of the $\Gamma_{\uparrow\uparrow}^{L}$
the injected electrons flip their spin providing a larger flux with
the spin that is opposite to the direction of the tip polarisation.
At the same time, the total current increases as the value of $\Gamma_{\uparrow\uparrow}^{L}$
gets bigger (this is in agreement with the right panel in Fig.\ref{fig:tip_up_FS}).

We can also observe from the left and middle panels of Fig. \ref{fig:tip_up_FS}
that both directions of the tip polarisation produce qualitatively
different spin splittings such that the difference $J_{\uparrow}-J_{\downarrow}$
of the spin-resolved currents is larger for $\mathscr{D}$ chirality
direction $\Downarrow$ ($\mathscr{L}$ chirality direction $\Uparrow$
, red curves) than for $\mathscr{D}$ chirality direction $\Uparrow$
($\mathscr{L}$ chirality direction $\Downarrow$, black curves).

Let us turn back to the total currents in the right panel of Fig.
\ref{fig:tip_up_FS}, which presents a zoomed voltage window for the
two distinct physical scenarios ($\mathscr{D}$$\Uparrow$ and $\mathscr{D}$$\Downarrow$)
for both regimes shown in the left two panels. We see that the largest
current is observed for the systems $\mathscr{D}$$\Downarrow$ and
$\mathscr{L}$$\Uparrow$ in regime A, while in regime B it is the
other way round: the largest current is observed for the systems $\mathscr{D}$$\Uparrow$
and $\mathscr{L}$$\Downarrow$ (see also the right panel in Fig.
\ref{fig:total}). It is this latter case which qualitatively agrees
with experimentally observed preference of the $\mathscr{D}$-helix
for up-emitted spin and the $\mathscr{L}$-helix for down \cite{aragones_measuring_2017}.
Therefore, regime A contradicts this particular set of experimental
data in the second order of correlations but agrees in the first.
As all the other parameters, barring the coupling to the leads, are
the same, the implication is that only electron-electron correlations
and spinterface effects influence whether spin flipping or filtering
is taking place.

The spin polarisation signal that is present at first and second order
of the perturbation theory - but entirely absent without interactions
- constitutes one of the central results of this work: electron-electron
correlations, even at the mean-field Hartree-Fock level, are qualitatively
necessary for the emergence of chirality- and magnetisation-dependent
spin selectivity in this system, and are not merely a perturbative
correction to an already spin-selective non-interacting baseline.
Moreover, moving beyond mean-field is necessary to find the qualitative
behaviour observed in the experiment: if in the first order the largest
currents are observed for the systems $\mathscr{D}$$\Downarrow$
and $\mathscr{L}$$\Uparrow$ which contradicts the experiment (see
the middle panel in Fig. \ref{fig:total} where regime B parameters
were used), in the second order with the inclusion of the dynamic
correlations the systems $\mathscr{D}$$\Uparrow$ and $\mathscr{L}$$\Downarrow$
appear to provide the largest current restoring the agreement with
the experiment \cite{aragones_measuring_2017} (the right panel in
the same Figure). 
\begin{figure}
\centering
\includegraphics[width=5cm]{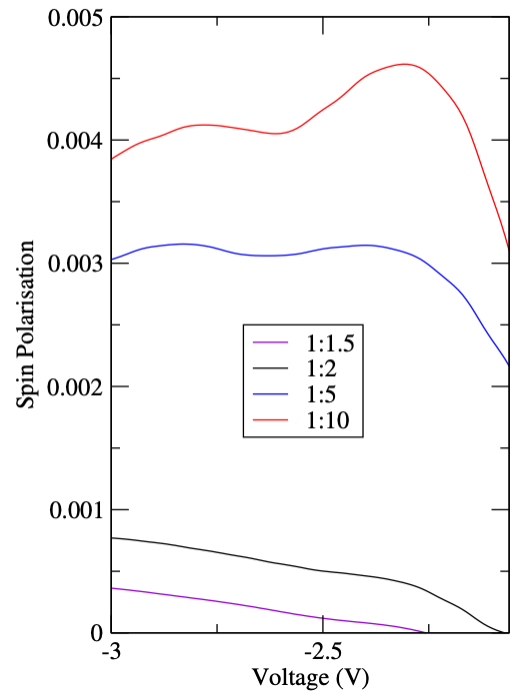}\caption{Spin polarisation factor Eq. (\ref{eq:SP-def}) for different spin-injection
ratios of 1:1.5, 1:2, 1:5 and 1:10, calculated for $\mathscr{D}$
chirality using the second order theory. Here the parameters are:
$\varepsilon=2.0$, $t=-0.4$, $\lambda=0.05$ and $U=0.8$, but for
the $\Gamma$-ratios of 1:1.5, 1:2, 1:5 and 1:10, we set $\Gamma_{\uparrow\uparrow}^{L}=1.5,2,5$
and $10$ and $\Gamma_{\downarrow\downarrow}^{L}=\Gamma^{R}=1$ for
the tip polarisation $\Uparrow$. }\label{fig:High_Gamma}
\end{figure}
Fig. \ref{fig:High_Gamma} shows the spin polarisation factor $\mathrm{SP}_{\mathscr{L}/\mathscr{D}}$\LyXZeroWidthSpace ,
computed via Eq. (\ref{eq:SP-def}), as a function of voltage within
the full range for four ratios, 1:1.5, 1:2, 1:5 and 1:10, of spin-up
to spin-down injection from the tip, $\Gamma_{\uparrow\uparrow}^{L}:\Gamma_{\downarrow\downarrow}^{L}$,
for $\mathscr{D}$. Increasing the spin polarisation ratio of the
injected electrons increases the magnitude of the SP signal across
the full voltage range, with the 1:10 ratio yielding the largest polarisation
for both chiralities. This is physically intuitive: a more strongly
spin-polarised source of electrons compounds with the helical spin
inverter to produce a larger detectable asymmetry; this is in line
with the experimental SP signal that depends sensitively on the degree
of the tip magnetisation. The SP curves exhibit a non-trivial voltage
dependence for all three ratios, with a pronounced peak structure
near $\approx-2.3$ V. At voltages $V\gtrsim-2$ V the spin polarisation
factor $\mathrm{SP}_{\mathscr{L}/\mathscr{D}}$ behaves erratically
remaining extremely small at the level of noise (not shown); we believe
this is due to small values of the spin currents and a very small
current spin polarisation in this voltage region resulting in substantial
numerical noise around the zero value (cf. Fig. \ref{fig:tip_up_FS}).
Notably, the $\mathscr{L}$ enantiomer displays exactly identical
SP curves, so we do not show them both. Importantly, however, the
overall magnitudes of SP obtained reach at most $\sim0.5$\% even
at the most favourable coupling ratio, implying that the substantially
increased coupling to the spin injection alone cannot account for
the sizeable polarisation of $\sim60$ \% observed experimentally
\cite{aragones_measuring_2017}.

\begin{figure*}
\centering
\includegraphics[totalheight=7cm]{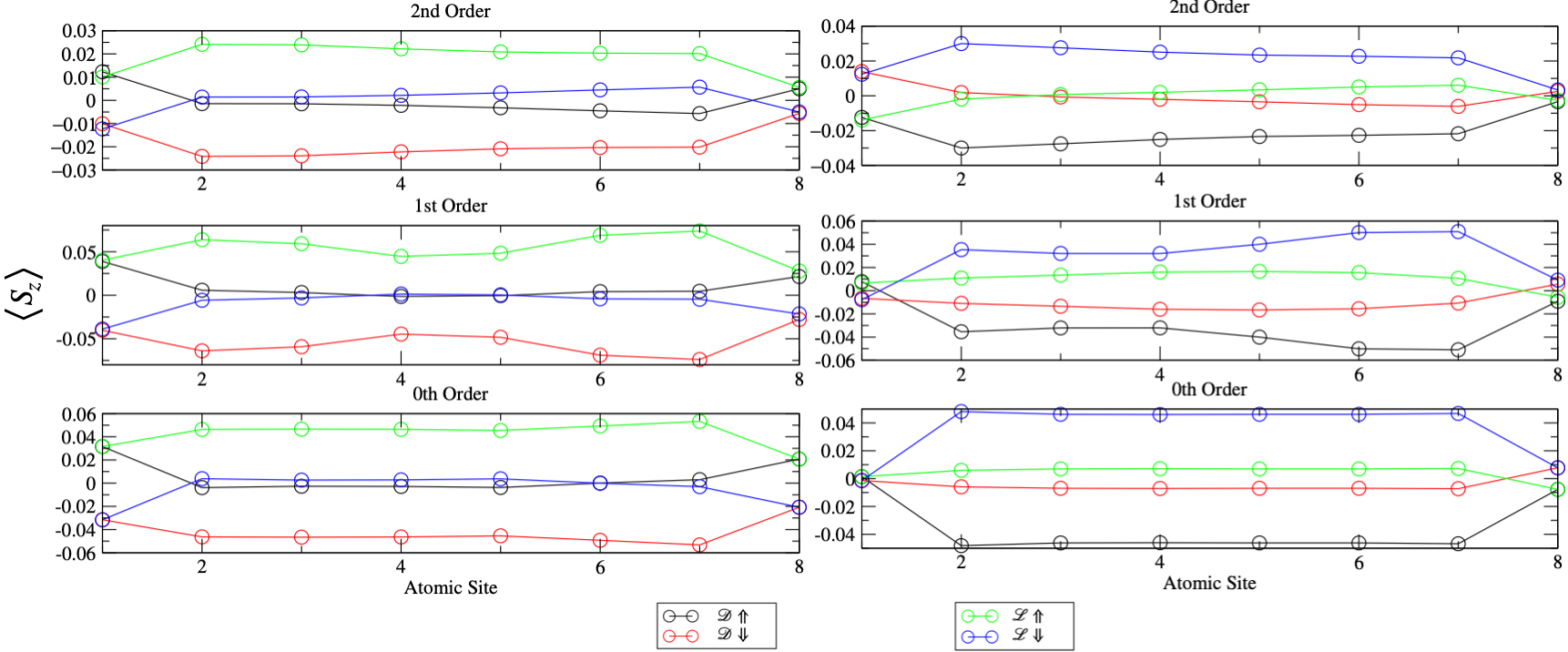}\caption{Site resolved spin expectation values $\left\langle S_{z}\right\rangle $
at -2.97 V bias for three scenarios: the 0th order (non-interacting)
case, and the first and second orders of electron correlations where
$U/t=2$, for all four system combinations. All the calculations were
run with the following parameters: $\varepsilon=2.0$, $t=-0.4$,
$\lambda=0.05$ and $U=0.8$, with the only difference being in the
level width functions. Left and right panels corresponds to regimes
A ($\Gamma_{\uparrow\uparrow}^{L}=0.6$ and $\Gamma_{\downarrow\downarrow}^{L}=\Gamma^{R}=0.4$)
and B ($\Gamma_{\uparrow\uparrow}^{L}=1.5$ and $\Gamma_{\downarrow\downarrow}^{L}=\Gamma^{R}=1.0$),
respectively, with the same values of the level width function as
in Fig. \ref{fig:total}. Site 1 is attached to the tip and site 8
to the surface.}\label{fig:spin_pop}
\end{figure*}
The site-resolved spin expectation values $\left\langle S_{z}\right\rangle $,
evaluated at $V=-2.97$ V across the three orders of perturbation
theory, provide microscopic insight into how the CISS effect manifests
locally along the chain and how it evolves with increasing orders
of interaction in both regimes, shown in Fig. \ref{fig:spin_pop}.
We first notice that the distribution of the spin across the helix
corresponds to the equivalence of the effective Hamiltonians mentioned
above: the spin distributions of systems $\mathscr{D}$$\Downarrow$
and $\mathscr{L}$$\Uparrow$, as well as of $\mathscr{D}$$\Uparrow$
and $\mathscr{L}$$\Downarrow$, are of opposite sign, as expected.
In regime A (left panels) when spin-up electrons are ejected (black
and green curves), the inner sites of the helical molecule for the
$\mathscr{D}$ chirality are almost zero, while for the $\mathscr{L}$
one the spin site polarisations are positive (meaning that spin-up
site populations are larger than spin-down ones). The reverse is true
when the spin-down electrons are injected (the blue and red curves).
In regime B when the spin-up or spin-down electrons are ejected into
the molecule, the site populations are opposite to those in regime
A: specifically, in $\mathscr{D}$$\Uparrow$ case (black) the spin
populations are negative, while for the $\mathscr{L}$$\Downarrow$
case (blue) they are positive. This is expected, as the electronic
spin flips when travelling through the molecule as was demonstrated
above.

The underlying spin selectivity builds up with successive inclusion
of electron correlations. If in regime A (left panels) the qualitative
distribution of the spin populations remain the same across all four
systems, it changes more noticeably in the case of regime B (right
panels): if in the 0th order the populations on inner sites of the
helix are the same, they become different in the first and second
orders, and they site dependence changes as going from the first to
the second order. 
\begin{figure}
\centering
\includegraphics[width=8cm]{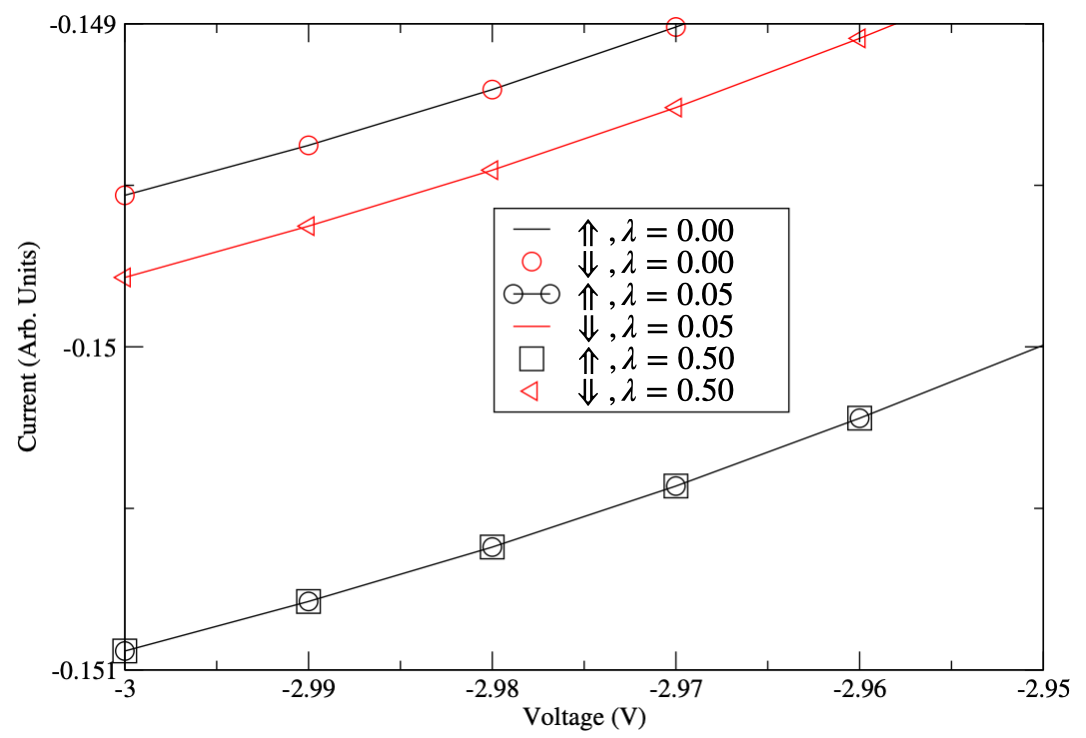}\caption{Total currents calculated for $U/t=2$ with $t=0.4$ for a spin injection
ratio of $1:10$. and three values of the SOC strength: $\lambda=0$,
$0.05$ and $0.5$. There is no discernible difference between the
low and high SOC cases.}\label{fig:high_soc_Gamma_currents}
\end{figure}
 As we have established that the overall magnitude remains minuscule
in comparison to the experiment, it is worth considering the behaviour
of the total currents with respect to the strength of the SOC, even
if we are to use physically unrealistic values for an organic $\alpha$-helix.
In Fig. \ref{fig:high_soc_Gamma_currents}, we see that at zero SOC
the currents for both tip polarisations remain the same. Increasing
the SOC tenfold from $\lambda=0.05$ to $\lambda=0.5$, we find that
there is not that much of a difference. We find that the SP is not
much larger than in Fig. \ref{fig:High_Gamma} and we just notice
that the higher SOC produces only a slightly larger current. This
implies that a stronger lead-molecule tunnelling is qualitatively
more important that the role of the SOC.

\section{Discussion and Conclusions}

Through this investigation, we find a clear spin polarisation of the
current: the currents for both spins differ. Moreover, we find that
the pattern in which both currents, $J_{\uparrow}$ and $J_{\downarrow}$,
appear, is different for the $\mathscr{L}$ and $\mathscr{D}$ chiralities
and for the same tip polarisation direction: if for one the two currents
span a wider window of values, for the other they spans a narrower
one, see Fig. \ref{fig:tip_up_FS}. The important point, however,
is that in experiment the currents for each spin cannot be separately
measured; only the total current, $J=J_{\uparrow}+J_{\downarrow}$,
is available. However, we have found that the difference in the total
currents is minimal for the two chiralities, contrary to the experimental
findings.

Next, we find that the total currents passing through both enantiomers
$\mathscr{D}$ and $\mathscr{L}$ indeed differ at the same tip polarisation
direction. This must be a clear manifestation of the CISS effect.
At the same time, the model that has been implemented does not allow
us to distinguish $\mathscr{D}\Uparrow$ and $\mathscr{L}\Downarrow$
systems, as well as $\mathscr{D}\Downarrow$ and $\mathscr{L}\Uparrow$
systems; hence, instead of observing four different behaviours for
four different systems ($\mathscr{D}\Uparrow$, $\mathscr{L}\Downarrow$,
$\mathscr{D}\Downarrow$ and $\mathscr{L}\Uparrow$), we have found
only two distinct situations ($\mathscr{D}\Uparrow$ and $\mathscr{L}\Uparrow$,
or, equivalently, $\mathscr{D}\Downarrow$ and $\mathscr{L}\Downarrow$).
This suggests that the experimentally observed asymmetry between enantiomers
in all four cases may be a sensitive probe of spinterface physics
at both ends of the junction - a feature our minimal model is not
designed to capture.

Two regimes were found. In regime A when the coupling to the leads
is of comparable magnitude to the hopping parameter $t$, the injected
electrons do not change their spin when propagating through the molecule,
and the dominant current flows in the same spin channel as the tip
polarisation. In regime B, when the coupling to the leads dominates
over $t$, the spin of the injected electrons is flipped upon traversing
the chiral molecule, and the dominant current flows in the opposite
spin channel to the tip polarisation. In the former regime the molecule
serves as a filter, in the latter - as an inverter.

We have only considered here calculations corresponding to a very
limited set of the parameters which, however, clearly show that there
are the two distinct spin transport regimes. It may be of interest
to perform a comprehensive exploration of the whole parameter set;
this is planned for a future study.

At the same time, the experiment sees the preference for the total
current being the largest for the systems $\mathscr{D}\Uparrow$ and
$\mathscr{L}\Downarrow$ \cite{aragones_measuring_2017}, and we indeed
find, in the second order of theory, qualitatively exactly the same
behaviour in regime B. We note, however, that when the level width
functions are chosen sufficiently small ensuring a weak molecule-leads
coupling (regime A), the opposite behaviour is observed in the second
order of theory, while the first order is agreeable.

We next find that SP monotonically increases with the ratio of injected
spins, but even in a highly favourable regime (Fig. \ref{fig:High_Gamma}),
the SP doesn't exceed $\sim$0.5\%. In spite of this, it is very clear
that the account for the electron-electron correlations is absolutely
necessary to observe CISS, evident in Fig. \ref{fig:total}, when
we considered the total currents in the four cases across the three
levels of theory.

Another striking takeaway here is the qualitative difference in the
behaviour of the current when going from the first order of interactions
to the second. We see a clear splitting in the total currents, where
the $\mathscr{L}$-enantiomer prefers spin up, while the $\mathscr{D}$-enantiomer
prefers spin down in the first order. However, when going into the
second order, where the exchange and bubble diagrams have a contribution,
the two chiralities favour the opposite spins: $\mathscr{L}$ favours
spin down and $\mathscr{D}$ favours spin up, which agrees with Aragones
\textit{et. al} \cite{aragones_measuring_2017}. Even though this
happens only in regime B, it indicates that in that regime the two
chiralities strongly affect the current spin polarisation. In that
sense, this study does confirm the existence of the CISS effect, even
though the magnitude of it, measured by the SP defined in Eq. (\ref{eq:SP-def}),
was found very small.

Clearly, we do observe the CISS effect; however, quantitatively the
results differ from experiment. Therefore, we must ask ourselves what
might be a reason (or reasons) for this discrepancy. This requires
careful discussion of all approximations that have been applied.

The persistent ceiling of $\sim$0.5\% of the SP factor, robust across
all values of $U/t$, $\lambda$, and tip polarisation ratio explored
here, may point to fundamental limitations of our model as a framework
for CISS. The first limitation concerns the model itself: we have
used the simplest possible many-electron Hamiltonian, which is the
Hubbard model. There are two successive approximations here: the neglect
of other terms in the full electron Hamiltonian and use of the single
orbital in the Hubbard terms. It is difficult to assess how the calculation
with the full Hamiltonian might change the picture as only the appropriate
calculations will tell, though this could not be done here due to
the sheer computational weight. Concerning the second point, with
one orbital per site and next-nearest-neighbour SOC, the system supports
only a single transport channel per spin, and while the Hubbard interaction
redistributes spectral weight between spin channels, it cannot generate
the rich quantum interference that arises from multiple co-propagating
pathways. It has been shown that a multi-orbital manifold per atomic
site allows for non-Abelian SU(2) interference across distinct transport
pathways that cannot be gauged away, producing substantially larger
SP factors even with modest SOC \cite{gutierrez_modelling_2013}.
Importantly, the differential broadening mechanism identified here
would be naturally amplified in a multi-orbital setting, where greater
spectral complexity would allow the second order diagrams to exploit
a richer spin-asymmetric structure in $G^{<}$ potentially yielding
SP factors much closer to those observed experimentally. Extending
the present framework to a multi-orbital Hubbard model, where each
site supports both $\sigma$ and $\pi$ orbitals as in a realistic
peptide backbone, may even remove the artificial symmetry degeneracy
noted in Eq. (\ref{eq:symmetry1}) and (\ref{eq:symmetry2}), whereby
instead of two unique cases we would find four individual scenarios
corresponding to each permutation of chirality and tip magnetisation,
which would then be much more akin to the experimental data.

The other point is related to using only four irreducible self-energy
diagrams in accounting for the electron correlation effects. Of course,
the second Born approximation accounts for an infinite number of reducible
self-energy diagrams built from these four irreducible insertions,
but these by all means are not all diagrams and therefore the results
may change if higher order irreducible diagrams are also included.
This observation is supported by our findings that by stepping from
the first to the second order when, alongside the Hartree and Fock
diagrams, two more irreducible insertions were added, the observed
picture changed qualitatively. At the same time, quantitatively, the
calculated SP did not change markedly, which might indicate that in
order to achieve experimentally observed 60\% polarisation, adding
more irreducible diagrams may not be enough.

This work demonstrates that electron-electron correlations beyond
the mean-field level are qualitatively necessary for the emergence
of chirality- and magnetisation-dependent spin selectivity in a helical
molecular junction. The second Born approximation reveals a competition
between static Hartree-Fock renormalisation and dynamical inelastic
broadening of spin-resolved quasiparticle peaks that is entirely absent
at lower orders of theory. This competition manifests as a non-trivial
inversion of chirality-spin preference, which we identify as an occupation-driven,
spectrally resolved effect which, depending on the strength of injection
from the tip, may force the molecule to acts either as a filter (weak
coupling) or an inverter (strong coupling). While the quantitative
discrepancy with experiment remains large, this study clarifies the
distinct and competing roles of static and dynamical electron correlations
in CISS, establishes lead-molecule coupling as the decisive parameter
governing which regime the system finds itself in, and identifies
multi-orbital physics as the ingredient most likely necessary to close
the gap with experiment- both by amplifying the differential broadening
mechanism through richer spectral structure and by opening the non-Abelian
interference pathways that a single-orbital model cannot support.

We hope that this study will useful to stimulate further CISS related
research.

\section*{Acknowledgements}

We would like to thank Rafael Gutierrez for many valuable discussions
and to Alejandro Santana Bonilla for his invaluable help in calculations
AK would like to express his gratitude to KCL and NMES for the funding
that made this research possible. \newpage{}

\section*{Appendix A: Single Orbital Hubbard Self-Energies}

\setcounter{equation}{0}
\renewcommand{\theequation}{A.\arabic{equation}}

\begin{figure}[H]
\centering
\centering{}\includegraphics[width=8cm]{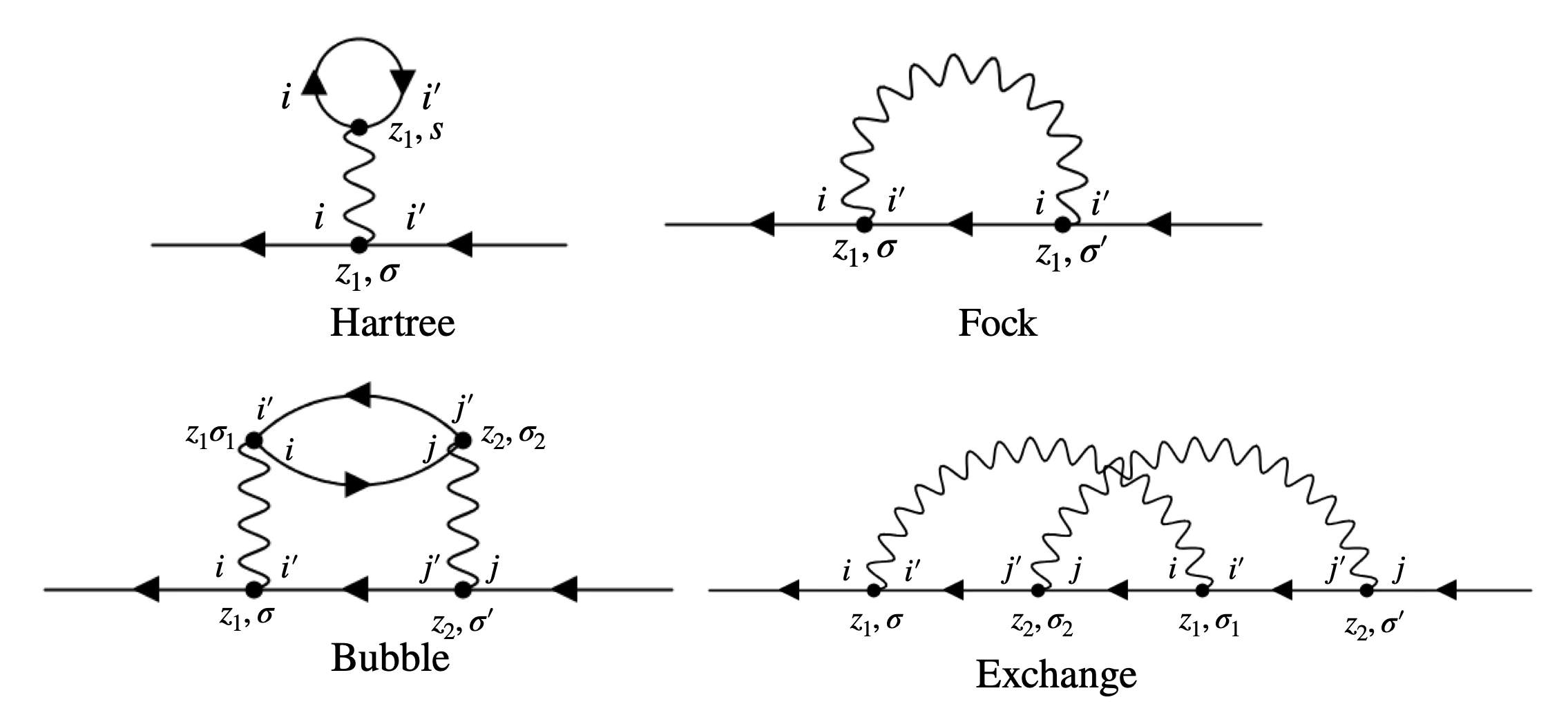}\caption{Hartree, Fock, Bubble and Exchange diagrams in the Hubbard model.}\label{fig:A_SBA_diagrams}
\end{figure}

In the first order, we only have the Hartree and Fock self-energy
diagrams shown in Fig. \ref{fig:A_SBA_diagrams}, for which we can
write, for the atomic Hubbard chain:

\begin{equation}
\Sigma_{i\sigma,i'\sigma'}^{H}\left(z_{1},z_{2}\right)=-i\hbar\delta_{z_{1}z_{2}}\delta_{ii'}\delta_{\sigma\sigma'}U_{i}\sum_{s}G_{i's,is}\left(z_{1},z_{1}^{+}\right),\label{eq:Spin_hub_hartree}
\end{equation}
\begin{equation}
\Sigma_{i\sigma,i'\sigma'}^{F}\left(z_{1},z_{2}\right)=i\hbar\delta_{z_{1}z_{2}}U_{i}G_{i'\sigma,i\sigma'}\left(z_{1},z_{1}^{+}\right).\label{eq:Spin_hub_fock}
\end{equation}
Then we project these expressions onto the retarded component, and
exploit their explicit dependence on the time difference to take their
Fourier transforms and obtain the Hartree and Fock self-energies in
omega space:

\begin{equation}
\Sigma_{i\sigma,i'\sigma'}^{r,H}\left(\omega\right)=-i\hbar\delta_{ii'}\delta_{\sigma\sigma'}U_{i}\sum_{s}G_{i's,is}^{<}\left(0\right),\label{eq:}
\end{equation}

\begin{equation}
\Sigma_{i\sigma,i'\sigma'}^{r,F}\left(\omega\right)=i\hbar U_{i}G_{i'\sigma,i\sigma'}^{<}\left(0\right).\label{eq:-2}
\end{equation}
In the first order we do not have any lesser projection by definition
as $z_{1}=z_{2}$. Here 
\[
G_{i\sigma,i\sigma}^{<}\left(0\right)=\int\frac{d\omega}{2\pi}G_{i\sigma,i\sigma}^{<}\left(\omega\right)\,.
\]

The second order is slightly less straightforward, but the same process
is valid. The Bubble and Exchange diagrams in Fig. \ref{fig:A_SBA_diagrams}
have the following self-energies:

\begin{widetext}

\begin{equation}
\Sigma_{i\sigma,j\sigma'}^{\mathrm{B}}\left(z_{1},z_{2}\right)=-\left(i\hbar\right)^{2}\delta_{ii'}\delta_{jj'}U_{i}U_{j}\sum_{\sigma_{1}\sigma_{2}}G_{i'\sigma_{1},j'\sigma_{2}}\left(z_{1},z_{2}\right)G_{j\sigma_{2},i\sigma_{1}}\left(z_{2},z_{1}\right)G_{i'\sigma,j'\sigma'}\left(z_{1},z_{2}\right),
\end{equation}
and
\begin{equation}
\Sigma_{i\sigma,j\sigma'}^{E}\left(z_{1},z_{2}\right)=\left(i\hbar\right)^{2}\delta_{ii'}\delta_{jj'}U_{i}U_{j}\sum_{\sigma_{1}\sigma_{2}}G_{i'\sigma,j'\sigma_{2}}\left(z_{1},z_{2}\right)G_{j\sigma_{2},i\sigma_{1}}\left(z_{2},z_{1}\right)G_{i'\sigma_{1},j'\sigma'}\left(z_{1},z_{2}\right).\label{eq:Spin_HUb_2b}
\end{equation}
We then project the self-energies onto the retarded and lesser components
and obtain the real time self-energies for the Bubble,
\begin{align}
\Sigma_{i\sigma,j\sigma'}^{r,B}\left(t_{1},t_{2}\right)= & -\left(i\hbar\right)^{2}\delta_{ii'}\delta_{jj'}U_{i}U_{j}\sum_{\sigma_{1}\sigma_{2}}\left[G_{i'\sigma_{1},j'\sigma_{2}}^{r}\right.\left(t_{1},t_{2}\right)G_{j\sigma_{2},i\sigma_{1}}^{<}\left(t_{2},t_{1}\right)G_{i'\sigma,j'\sigma'}^{>}\left(t_{1},t_{2}\right)\nonumber \\
 & +G_{i'\sigma_{1},j'\sigma_{2}}^{<}\left(t_{1},t_{2}\right)G_{j\sigma_{2},i\sigma_{1}}^{<}\left(t_{2},t_{1}\right)G_{i'\sigma,j'\sigma'}^{r}\left(t_{1},t_{2}\right)\nonumber \\
 & +G_{i'\sigma_{1},j'\sigma_{2}}^{<}\left(t_{1},t_{2}\right)G_{j\sigma_{2},i\sigma_{1}}^{a}\left(t_{2},t_{1}\right)\left.G_{i'\sigma,j'\sigma'}^{<}\left(t_{1},t_{2}\right)\right],
\end{align}
\begin{equation}
\Sigma_{i\sigma,j\sigma'}^{<,B}\left(t_{2},t_{1}\right)=-\left(i\hbar\right)^{2}\delta_{ii'}\delta_{jj'}U_{i}U_{j}\sum_{\sigma_{1}\sigma_{2}}G_{i'\sigma_{1},j'\sigma_{2}}^{<}\left(t_{1},t_{2}\right)G_{j\sigma_{2},i\sigma_{1}}^{>}\left(t_{1},t_{2}\right)G_{i'\sigma,j'\sigma'}^{<}\left(t_{1},t_{2}\right),
\end{equation}
and Exchange diagrams,
\begin{align}
\Sigma_{i\sigma,j\sigma'}^{r,E}\left(t_{2},t_{1}\right)= & \left(i\hbar\right)^{2}\delta_{ii'}\delta_{jj'}U_{i}U_{j}\sum_{\sigma_{1}\sigma_{2}}\left[G_{i'\sigma,j'\sigma_{2}}^{r}\left(t_{1},t_{2}\right)\right.G_{j\sigma_{2},i\sigma_{1}}^{<}\left(t_{2},t_{1}\right)G_{i'\sigma_{1},j'\sigma'}^{>}\left(t_{1},t_{2}\right)\nonumber \\
 & +G_{i'\sigma,j'\sigma_{2}}^{<}\left(t_{1},t_{2}\right)G_{j\sigma_{2},i\sigma_{1}}^{<}\left(t_{2},t_{1}\right)G_{i'\sigma_{1},j'\sigma'}^{r}\left(t_{1},t_{2}\right)\nonumber \\
 & +G_{i'\sigma,j'\sigma_{2}}^{<}\left(t_{1},t_{2}\right)G_{j\sigma_{2},i\sigma_{1}}^{a}\left(t_{2},t_{1}\right)\left.G_{i'\sigma_{1},j'\sigma'}^{<}\left(t_{1},t_{2}\right)\right],
\end{align}
\begin{equation}
\Sigma_{i\sigma,j\sigma'}^{<,E}\left(t_{1},t_{2}\right)=\left(i\hbar\right)^{2}\delta_{ii'}\delta_{jj'}U_{i}U_{j}\sum_{\sigma_{1}\sigma_{2}}G_{i'\sigma,j'\sigma_{2}}^{<}\left(t_{1},t_{2}\right)G_{j\sigma_{2},i\sigma_{1}}^{>}\left(t_{2},t_{1}\right)G_{i'\sigma_{1},j'\sigma'}^{<}\left(t_{1},t_{2}\right).
\end{equation}
These can easily be seen to depend on the times difference, $t_{2}-t_{1}$.
Therefore, using the Fourier transform, we obtain the four self-energy
projections of the second-order diagrams in the frequency space:

\begin{align}
\Sigma_{i\sigma,j\sigma'}^{r,B}\left(\omega\right)= & -\left(i\hbar\right)^{2}\delta_{ii'}\delta_{jj'}U_{i}U_{j}\sum_{\sigma_{1}\sigma_{2}}\int\frac{d\omega_{1}d\omega_{2}}{\left(2\pi\right)^{2}}\left[G_{i'\sigma_{1},j'\sigma_{2}}^{r}\left(\omega_{1}\right)G_{j\sigma_{2},i\sigma_{1}}^{<}\left(\omega_{2}\right)G_{i'\sigma,j'\sigma'}^{>}\left(\omega-\omega_{1}+\omega_{2}\right)\right.\nonumber \\
 & +G_{i'\sigma_{1},j'\sigma_{2}}^{<}\left(\omega_{1}\right)G_{j\sigma_{2},i\sigma_{1}}^{<}\left(\omega_{2}\right)G_{i'\sigma,j'\sigma'}^{r}\left(\omega-\omega_{1}+\omega_{2}\right)\nonumber \\
 & +G_{i'\sigma_{1},j'\sigma_{2}}^{<}\left(\omega_{1}\right)G_{j\sigma_{2},i\sigma_{1}}^{a}\left(\omega_{2}\right)\left.G_{i'\sigma,j'\sigma'}^{<}\left(\omega-\omega_{1}+\omega_{2}\right)\right],
\end{align}
\begin{equation}
\Sigma_{i\sigma,j\sigma'}^{<,B}\left(\omega\right)=-\left(i\hbar\right)^{2}\delta_{ii'}\delta_{jj'}U_{i}U_{j}\sum_{\sigma_{1}\sigma_{2}}\int\frac{d\omega_{1}d\omega_{2}}{\left(2\pi\right)^{2}}G_{i'\sigma_{1},j'\sigma_{2}}^{<}\left(\omega_{1}\right)G_{j\sigma_{2},i\sigma_{1}}^{>}\left(\omega_{2}\right)G_{i'\sigma,j'\sigma'}^{<}\left(\omega-\omega_{1}+\omega_{2}\right),
\end{equation}

\begin{align}
\Sigma_{i\sigma,j\sigma'}^{r,E}\left(\omega\right)= & \left(i\hbar\right)^{2}\delta_{ii'}\delta_{jj'}U_{i}U_{j}\sum_{\sigma_{1}\sigma_{2}}\int\frac{d\omega_{1}d\omega_{2}}{\left(2\pi\right)^{2}}\left[G_{i'\sigma,j'\sigma_{2}}^{r}\left(\omega_{1}\right)\right.G_{j\sigma_{2},i\sigma_{1}}^{<}\left(\omega_{2}\right)G_{i'\sigma_{1},j'\sigma'}^{>}\left(\omega-\omega_{1}+\omega_{2}\right)\nonumber \\
 & +G_{i'\sigma,j'\sigma_{2}}^{<}\left(\omega_{1}\right)G_{j\sigma_{2},i\sigma_{1}}^{<}\left(\omega_{2}\right)G_{i'\sigma_{1},j'\sigma'}^{r}\left(\omega-\omega_{1}+\omega_{2}\right)\nonumber \\
 & +G_{i'\sigma,j'\sigma_{2}}^{<}\left(\omega_{1}\right)G_{j\sigma_{2},i\sigma_{1}}^{a}\left(\omega_{2}\right)\left.G_{i'\sigma_{1},j'\sigma'}^{<}\left(\omega-\omega_{1}+\omega_{2}\right)\right],
\end{align}
\begin{equation}
\Sigma_{i\sigma,j\sigma'}^{<,E}\left(\omega\right)=\left(i\hbar\right)^{2}\delta_{ii'}\delta_{jj'}U_{i}U_{j}\sum_{\sigma_{1}\sigma_{2}}\int\frac{d\omega_{1}d\omega_{2}}{\left(2\pi\right)^{2}}G_{i'\sigma,j'\sigma_{2}}^{<}\left(\omega_{1}\right)G_{j\sigma_{2},i\sigma_{1}}^{>}\left(\omega_{2}\right)G_{i'\sigma_{1},j'\sigma'}^{<}\left(\omega-\omega_{1}+\omega_{2}\right).
\end{equation}
Using explicit expressions for all components of the self-energies,
it is easy to verify that \textit{at equilibrium} (i.e, when the bias
is zero) the four components (lesser, greater, retarded and advanced)
are related to each other in full agreement with the fluctuation-dissipation
theorem \cite{stefanucci_nonequilibrium_2013}. These expressions
guarantee that at zero bias the current is zero, as expected.

\end{widetext}

\newpage{}

\bibliographystyle{unsrt}
\addcontentsline{toc}{section}{\refname}\bibliography{Citations}

\end{document}